\documentclass[useAMS,usenatbib,a4paper]{mn2e}

\usepackage{amssymb,amsmath}
\usepackage{graphicx}
\usepackage{natbib}
\usepackage{aas_macros}

\newcommand{\hii}{H~{\sc ii}}

\newcommand{\nel}{n_\mathrm{e}}
\newcommand{\nion}{n_\mathrm{i}}
\newcommand{\kb}{k_\mathrm{B}}

\newcommand{\mpr}{m_\mathrm{p}}
\newcommand{\me}{m_\mathrm{e}}
\newcommand{\e}{\mathrm{e}}
\newcommand{\cs}{c_\mathrm{s}}

\begin{document}

\title[Understanding hydrogen recombination line observations with ALMA and EVLA]{Understanding hydrogen recombination line observations with ALMA and EVLA}

\author[Thomas Peters, Steven N. Longmore and Cornelis P. Dullemond]{
Thomas Peters$^{1,2,3}$\thanks{E-Mail: tpeters@physik.uzh.ch}, Steven N. Longmore$^4$ and Cornelis P. Dullemond$^1$\vspace{0.4cm}\\
\parbox{\textwidth}{$^1$Zentrum f\"{u}r Astronomie der Universit\"{a}t Heidelberg,
Institut f\"{u}r Theoretische Astrophysik, Albert-Ueberle-Str. 2,
D-69120 Heidelberg, Germany\\
$^2$Fellow of the Baden-W\"{u}rttemberg Stiftung\\
$^3$Institut f\"{u}r Theoretische Physik, Universit\"{a}t Z\"{u}rich,
Winterthurerstrasse 190, CH-8057 Z\"{u}rich, Switzerland\\
$^4$European Southern Observatory, Karl-Schwarzschild-Str. 2, D-85748 Garching bei M\"{u}nchen, Germany}}

\maketitle

\begin{abstract}
Hydrogen recombination lines are one of the major diagnostics of
\hii\ region physical properties and kinematics.
In the near future, the Expanded Very Large Array (EVLA) and the
Atacama Large Millimeter Array (ALMA) will allow observers to study
recombination lines in the radio and sub-mm regime in unprecedented detail.
In this paper, we study the properties of recombination
lines, in particular at ALMA wavelengths. We find that such lines will lie in almost
every wideband ALMA setup and that the line emission will be equally detectable
in all bands. Furthermore, we present our implementation of hydrogen recombination
lines in the adaptive-mesh radiative transfer code RADMC-3D. We
particularly emphasize the importance of non-LTE (local
thermodynamical equilibrium) modeling since non-LTE effects can
drastically affect the line shapes and produce asymmetric line profiles
from radially symmetric \hii\ regions. We demonstrate how these non-LTE
effects can be used as a probe of systematic motions (infall \& outflow) in the gas.
We use RADMC-3D to produce synthetic
observations of model \hii\ regions and study the necessary conditions
for observing such asymmetric line profiles with ALMA and EVLA.
\end{abstract}

\section{Introduction}

Massive stars influence their cosmic environment through powerful
winds, radiation feedback and supernova explosions.
Ionizing radiation from massive stars produces pronounced
\hii\ regions around them. Observations of \hii\ regions, while still in the ultracompact phase (with
diameters less than $0.1$~pc), provide important
insight into massive star formation
\citep{habis79,churchwell02,zinnyork07}. Observations of hydrogen
recombination lines at radio and sub-mm frequencies (RRLs) are routinely used by observers to infer
densities, temperatures and velocity structures inside \hii\ regions
\citep{gorsor02}.

While the microphysics of recombination lines was reasonably well
understood relatively early \citep{dupreegold70}, the \hii\ regions
themselves had to be modeled either on scales larger than the scale of
gravitational collapse using numerical radiation-hydrodynamics
\citep[see reviews by][]{yorke86,maclow07,klessenetal11} or on smaller
scales with simple spherically \citep{brownetal78,keto02,keto03} or
non-spherically \citep{keto07} symmetric models. Only recently,
collapse simulations of massive star formation with ionization
feedback have left the constriction of two dimensions
\citep{richyork97} and facilitated fully three-dimensional dynamical
simulations of \hii\ region expansion during massive star formation
\citep{petersetal10a,petersetal10b,petersetal10c,petersetal11a}.

These collapse simulations require adaptive-mesh simulations with 10
refinement levels and more. To post-process such data, dedicated
radiative transfer tools must be developed. To create synthetic
recombination line observations of the simulated \hii\ regions, we
have implemented recombination lines in RADMC-3D. RADMC-3D is a
radiative transfer code that can handle both continuum
\citep[e.g.][]{petersetal10b} and line \citep[e.g.][]{shettyetal11}
radiation on arbitrary octree meshes. It has an interface for PARAMESH
\citep{macneiceetal00}, the grid library of the adaptive-mesh code
FLASH \citep{fryxell00}, which was used for these simulations. We have
used RADMC-3D previously to model free-free and dust continuum
emission from the simulated \hii\ regions \citep{petersetal10b}.

To the best of the authors' knowledge, the only other codes that can
model recombination line observations are MOLLIE \citep{keto90}, which
we have used in previous work to post-process simulation data \citep{petersetal10a}
and for radiative transfer modeling \citep{longmoreetal11},
and an innominate code that \citet{mapietal93}
used for recombination line models of MWC~349.
The advantages of RADMC-3D are its direct compatibility with many major
simulation codes and its modularity.

Another important feature of RADMC-3D is the possibility to create
user-defined setups. Thus, it can be used by observers interested in
creating simple analytic models of the regions they are observing. In
the present paper, we describe the implementation of hydrogen
recombination lines in RADMC-3D and demonstrate its capability with
synthetic observations of several \hii\ region models. The development
of such a tool is particularly timely because of
the commissioning of the Expanded Very Large Array (EVLA) and the
Atacama Large Millimeter Array (ALMA). These facilities are set to
open new frontiers in terms of sensitivity, angular resolution,
dynamic range and image fidelity in the cm to sub-mm wavelength
regimes. Tools that can help to interpret such new observational data
are clearly highly desirable.

The purpose of the present paper is twofold. First, we present
our implementation of hydrogen recombination lines in RADMC-3D
and describe a suite of tests of this implementation. The user-defined
analytical setups are ideal for code validation before
applying the method to the much more complex simulation data. Second,
we use these analytical models to simulate synthetic EVLA and ALMA
observations. We particularly discuss the appearance of line
asymmetries expected for ALMA observations.

The paper is organized as follows. Section~\ref{sec:physrecomb}
briefly summarizes the physics of hydrogen recombination lines.
In Section~\ref{sec:impl}, we describe the implementation of recombination lines in
RADMC-3D. In Section~\ref{sec:rrl_properties}, we discuss
the general properties of RRL transitions at ALMA frequencies.
We then focus on the observation that the RRL profiles can
be asymmetric (Section~\ref{sec:asymm_profiles}) and investigate
the conditions under which we can expect to observe such asymmetries
(Section~\ref{sec:preditc_conditions}). We conclude our paper with
a summary of our results in Section~\ref{sec:summary}.
Additional information on line profile asymmetries can be
found in Appendix~\ref{sec:symm}. We present
our code tests in Appendix~\ref{sec:tests}.

\section{Physics of Hydrogen Recombination Lines}
\label{sec:physrecomb}

We start with briefly summarizing the physics of free-free continuum
radiation and hydrogen recombination lines in the radio and
sub-millimeter regime (see \citet{gorsor02} for more details). The hot
plasma in an \hii\ region gives rise to the emission of thermal
bremsstrahlung. This free-free radiation causes a continuum opacity\footnote{
In the following, the subscript $\mathrm{C}$ stands for ``continuum'', whereas the subscript $\mathrm{L}$ means ``line''.}
at frequency $\nu$ of
\begin{align}
\alpha_{\nu,\mathrm{C}} = 0.212\left(\frac{\nel}{1\mbox{ cm}^{-3}}\right) \left(\frac{\nion}{1\mbox{ cm}^{-3}}\right)
&\left(\frac{T}{1\mbox{  K}}\right)^{-1.35} \\
&\times \left(\frac{\nu}{1\mbox{ Hz}}\right)^{-2.1} \mathrm{cm^{-1}} \nonumber
\end{align}
with the electron and ion number densities $\nel$ and $\nion$, respectively, and the gas temperature $T$.
Likewise, the plasma emits radiation with an emissivity of
\begin{equation}
j_{\nu,\mathrm{C}} = B_\nu(T) \alpha_{\nu,\mathrm{C}} ,
\end{equation}
where $B_\nu(T)$ denotes the intensity of a blackbody of temperature $T$ at frequency $\nu$.

During recombination, an excited hydrogen atom emits recombination line radiation when the electron falls from
at state with principal quantum number $m$ to a state with principal quantum number $n$.
The corresponding line absorption coefficient is
\begin{equation}
\label{eq:alplinegen}
\alpha_{\nu,_\mathrm{L}} = \frac{h \nu}{4 \pi} \phi_\nu (N_{n} B_{n,m} - N_{m} B_{m,n})
\end{equation}
with the Planck constant $h$, the line profile function $\phi_\nu$, the Einstein coefficients $B_{n,m}$ and $B_{m,n}$
for absorption and stimulated emission, respectively, and the number densities of atoms $N_{k}$ in state $k$ (for $k = m, n$).
The line emissivity is
\begin{equation}
\label{eq:jldef}
j_{\nu,\mathrm{L}} = \frac{h \nu}{4 \pi} \phi_\nu N_{m} A_{m,n}
\end{equation}
with the Einstein coefficient for spontaneous emission
\begin{equation}
A_{m,n} = \frac{2 h \nu^3}{c^2} B_{m,n}
\end{equation}
with the speed of light $c$.

The Einstein coefficients satisfy the relation
\begin{equation}
g_l B_{l,k} = g_k B_{k,l}
\end{equation}
with the statistical weights $g_k = 2 k^2$. They are usually expressed in terms of the oscillator strengths
\begin{equation}
f_{n,m} = \frac{\me c h \nu}{4 \pi^2 \e^2} B_{n,m}
\end{equation}
with the electron mass $\me$ and the electron charge $\e$.
The oscillator strengths for hydrogen can be approximated for large $n$ and small $\Delta n = m - n$ following \citet{menzel68} as
\begin{equation}
f_{n,m} \approx n M_{\Delta n} \left(1 + 1.5 \frac{\Delta n}{n}\right)
\end{equation}
with $M_{\Delta n} = 0.190775$, $0.026332$, $0.0081056$, $0.0034917$, $0.0018119$, $0.0010585$ for $\Delta n = 1, 2, 3, 4, 5, 6$, respectively.

The situation is more complicated for recombination lines in the ALMA bands because lines above 100\,GHz
show non-negligible fine-structure splitting \citep{towleetal96}. For each principle quantum number $k$ there
are $k$ different energies which depend on the total angular momentum $j = 1/2, 3/2, \ldots, k - 1/2$. These $k$ energies
have a degeneracy
\begin{equation}
g_{k,j} = 
\begin{cases}
2 (2 j + 1) & j \leq k - 3/2\\
2j + 1 & j = k - 1/2.
\end{cases}
\end{equation}
The fine-structure components typically span a frequency range of the order MHz and thus cannot individually be resolved
with ALMA due to their thermal broadening in the ionized gas, which can be much larger than their separation. However,
the splitting can change the line profile, and the frequencies and Einstein coefficients must be computed
using relativistic quantum mechanics \citep{towleetal96}.

The line profile function $\phi_\nu$ is in general a convolution of a Gaussian profile $\phi_{\nu}^{\mathrm{G}}$ caused by
thermal and microturbulent broadening and a Lorentzian profile $\phi_{\nu}^{\mathrm{L}}$ caused by electron pressure broadening \citep{brockseat72}.
The Gaussian profile is given by
\begin{equation}
\phi_{\nu}^{\mathrm{G}} = \frac{1}{\sqrt{2 \pi} \sigma} \exp\left[-\frac{(\nu - \nu_0)^2}{2 \sigma^2}\right]
\end{equation}
with
\begin{equation}
\label{eq:sigma}
\sigma^2 = \frac{\nu_0^2}{2 c^2} \left(\frac{2 \kb T}{\mpr} + \xi^2\right) ,
\end{equation}
where $\mpr$ is the proton mass, $\kb$ the Boltzmann constant, $\nu_0$ the rest frequency of the line
and $\xi$ the root-mean-square of the microturbulent velocity field.
The Lorentzian profile can be written as
\begin{equation}
\phi_{\nu}^{\mathrm{L}} = \frac{\delta}{\pi} \frac{1}{(\nu - \nu_0)^2 + \delta^2} .
\end{equation}
The scale parameter $\delta$ can be approximated following \citet{brocklee71} as
\begin{equation}
\label{eq:delta}
\delta = 4.7 \left(\frac{n}{100}\right)^{4.4} \left(\frac{T}{10^4 \mbox{ K}}\right)^{-0.1}
\left(\frac{\nel}{1\,\mathrm{cm}^{-3}}\right) \mathrm{Hz} .
\end{equation}
The convolution of these two profiles then gives a Voigt profile \citep{rybickietal79}
\begin{equation}
\label{eq:voigtprof}
\phi_\nu = \frac{1}{\sqrt{2 \pi} \sigma} H(a, x)
\end{equation}
with $a = \delta / \sqrt{2} \sigma$, $x = (\nu - \nu_0) / \sqrt{2} \sigma$ and the Voigt function
\begin{equation}
\label{eq:voigtfun}
H(a, x) = \frac{a}{\pi} \int_{-\infty}^\infty \frac{\exp(-t^2)\,\mathrm{d}t}{a^2 + (t - x)^2} .
\end{equation}

To determine the occupation numbers, we first act on the assumption of local thermodynamic
equilibrium (LTE) and then consider possible departures from the LTE occupation numbers.
Under the assumption of LTE, the number density of hydrogen atoms in state $k$ 
with energy $E_{k}$ is given by the Saha-Boltzmann equation
\begin{equation}
N_{k}^{\mathrm{LTE}} = \frac{\nel \nion}{T^{3 / 2}} \frac{k^2 h^3}{(2 \pi m \kb)^{3 / 2}}
\exp\left(\frac{E_{k}}{\kb T}\right) .
\end{equation}
Here, we follow the the sign convention of \citet{gorsor02}, where $E_k > 0$, and the factor
$k^2$ derives from the statistical weight $g_k$. See \citet{osterbrock89} for a derivation of this equation.
With these occupation number densities, we can evaluate the absorption coefficient~\eqref{eq:alplinegen}
and the emissivity
\begin{equation}
j_{\nu,\mathrm{L}}^{\mathrm{LTE}} = B_\nu(T) \alpha_{\nu,\mathrm{L}}^{\mathrm{LTE}} .
\end{equation}
In general, the occupation numbers will be different from the LTE values. This deviation can be
measured with a departure coefficient
\begin{equation}
\label{eq:bk}
N_{k} = b_{k} N_{k}^{\mathrm{LTE}} .
\end{equation}
The coefficient $b_k$ represents the fractional departure of $N_{k}$ from $N_{k}^{\mathrm{LTE}}$, with $b_k=1$ meaning
no deviation at all.
From equation~\eqref{eq:jldef} it follows that the emissivity under non-LTE conditions is
\begin{equation}
j_{\nu,\mathrm{L}} = b_m j_{\nu,\mathrm{L}}^{\mathrm{LTE}} ,
\end{equation}
whereas the absorption coefficient~\eqref{eq:alplinegen} is usually written in the form
\begin{equation}
\alpha_{\nu,\mathrm{L}} = b_n \beta_{n,m} \alpha_{\nu,\mathrm{L}}^{\mathrm{LTE}}
\end{equation}
with
\begin{equation}
\beta_{n,m} = \frac{\displaystyle 1 - \frac{b_{m}}{b_{n}}
\exp\left(-\frac{h \nu}{\kb T}\right)}{\displaystyle 1 - \exp\left(-\frac{h \nu}{\kb T}\right)} .
\end{equation}
In the case of fine-structure splitting, we assume that the $g_k$ states $g_{k,j}$ that correspond
to principal quantum number $k$ are equally occupied\footnote{We could also assume that the fine-structure
levels follow a Boltzmann distribution. However, the fine-structure splitting changes the energy only by a very
small fraction and the difference in the energy levels is weak compared to the variation of the corresponding
Einstein coefficients. Hence, given that all fine-structure components overlap due to thermal broadening, we do
not expect to see noticeable differences in this case.}.
Thus, the occupation number for the fine-structure component with total angular momentum $j$ is
$N_{k,j} = g_{k,j} N_{k} / g_k$.

The radiative transfer problem for recombination lines is commonly solved under the approximation
that Thompson scattering can be neglected. We thus arrive at the radiative transfer equation
\begin{equation}
\label{eq:rt}
\frac{\mathrm{d} I_\nu}{\mathrm{d} s} = (j_{\nu,\mathrm{C}} + j_{\nu,\mathrm{L}}) -
(\alpha_{\nu,\mathrm{C}} + \alpha_{\nu,\mathrm{L}}) I_\nu
\end{equation}
for the intensity $I_\nu$ at frequency $\nu$. With the optical depth
\begin{equation}
\tau_\nu(r) = \int_0^r (\alpha_{\nu,\mathrm{C}}(r') + \alpha_{\nu,\mathrm{L}}(r'))\,\mathrm{d}r'
\end{equation}
and the source function
\begin{equation}
\label{eq:sourfun}
S_\nu = \frac{j_{\nu,\mathrm{C}} + j_{\nu,\mathrm{L}}}{\alpha_{\nu,\mathrm{C}} + \alpha_{\nu,\mathrm{L}}}
\end{equation}
Equation~\eqref{eq:rt} can be written as
\begin{equation}
\label{eq:rtint}
I_\nu(\tau_\nu) = \mathrm{e}^{-\tau_\nu} \int_0^{\tau_\nu} \mathrm{e}^{\tau_\nu'} S_\nu(\tau_\nu')\,\mathrm{d}\tau_\nu' ,
\end{equation}
where we have neglected background irradiation.

%------------------------------------------------------------
\section{Implementation}
\label{sec:impl}

We have described our implementation of free-free radiation in RADMC-3D earlier \citep{petersetal10b},
so that we focus here on the implementation of recombination lines. The Voigt function~\eqref{eq:voigtfun}
is calculated with the source code provided by \citet{schreier92}. The algorithm is an improved version of the
rational approximation method originally developed by \citet{humlicek82}.

The departure coefficients $b_k$ for the occupation number density $N_k$ of quantum number $k$
are pre-calculated in tabulated form with a program published in \citet{gorsor02}.
Since the coefficients vary smoothly with $\log T$ and $\log \nel$, they can be interpolated bi-linearly from the
tabulated values during the raytracing. The algorithm to calculate the $b_k$ is based on the matrix condensation technique
that was originally presented in \citet{brocklehurst70} and \citet{brocksal77} and later
extended by \citet{walmsley90} towards smaller quantum numbers. In addition to the dependence
on $T$ and $\nel$, $b_k$ also depends on the ambient radiation field \citep[see][]{brocksal77}. The radiation
field is characterized by the radiation temperature ($10^4$\,K) and the emission measure ($10^6\,$pc\,cm$^{-6}$).

\begin{figure}
\centerline{\includegraphics[height=170pt]{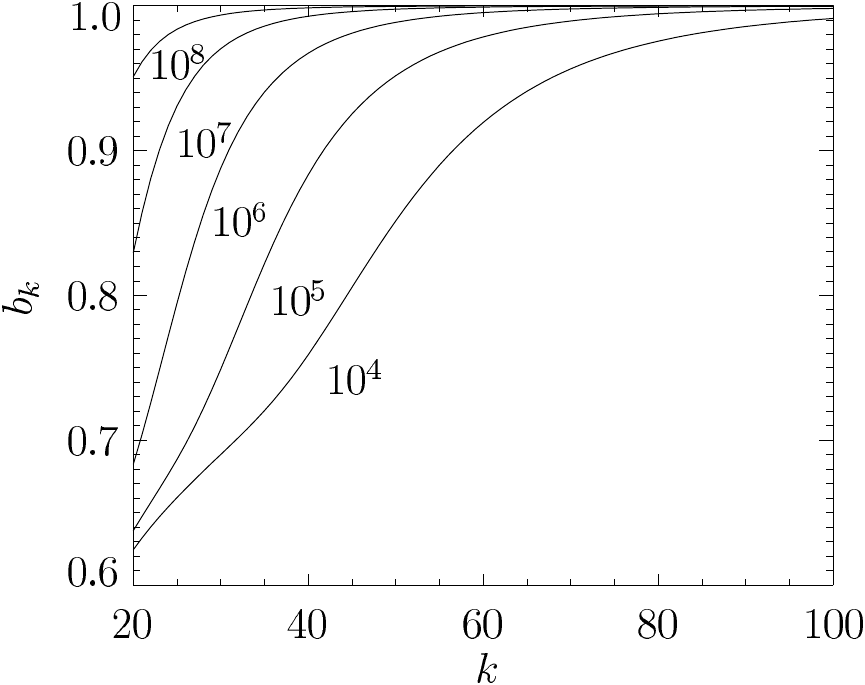}}
\caption{Departure coefficient $b_k$ as function of the principal quantum number $k$ at temperature $T = 10^4\,$K
for electron number densities $\nel$ between $10^4$ and $10^8$\,cm$^{-3}$.
Compare Figure 1 in \citet{walmsley90}.}
\label{fig:walms1}
\end{figure}

\begin{figure}
\centerline{\includegraphics[height=170pt]{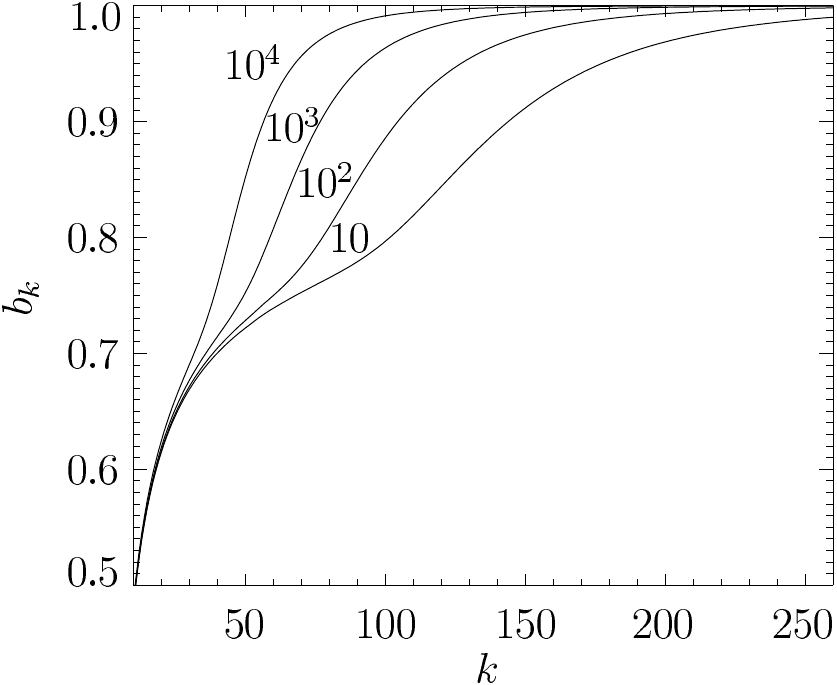}}
\caption{Departure coefficient $b_k$ as function of the principal quantum number $k$ at temperature $T = 10^4\,$K
for electron number densities $\nel$ between $10$ and $10^4$\,cm$^{-3}$. Compare Figure 2 in \citet{sejnhjell69}.}
\label{fig:sejnhjell}
\end{figure}

We have verified the program by reproducing some of the numerical values tabulated in \citet{brocklehurst70},
\citet{brocksal77}, \citet{salbrock79} and \citet{walmsley90}. As an example, we present in Figure~\ref{fig:walms1}
the variation of $b_k$ with principal quantum number $k$ at $T = 10^4\,$K for different electron
number densities $\nel$ ranging from $10^4$ to $10^8$\,cm$^{-3}$. The resulting plot is identical to
Figure~1 in \citet{walmsley90}. To furthermore estimate the variation of the results obtained
from different methods, we compare the $b_k$ numbers to values that \citet{sejnhjell69} computed independently.
Figure~\ref{fig:sejnhjell} shows $b_k$ as function of $k$ at $T = 10^4\,$K for $\nel$ between $10$ and $10^4$\,cm$^{-3}$.
The deviation from Figure~2 in \citet{sejnhjell69} is as small as expected given the different method
of computing the departure coefficients, but according to \citet{gorsor02} the \citet{walmsley90} values are more accurate.

\citet{gorsor02} argue that the departure coefficients calculated by \citet{stohum95} are the most
precise ones at the smallest quantum numbers, where deviations between the different calculations occur.
However, the discrepancy between the \citet{walmsley90} and \citet{stohum95}
coefficients is less than 10\% in the case they discuss. Given this relatively small difference, we will work in this paper with the
\citet{walmsley90} coefficients but plan to use more accurate numbers in future work.

To calculate the fine-structure splitting of hydrogen, we have used the code presented in \citet{gorsor02}.
We have verified the implementation by reproducing the values in Table~3 of \citet{towleetal96} for the
frequencies and intensities of some recombination lines.

We refer to Appendix~\ref{sec:tests} for a presentation of our code verification.

%------------------------------------------------------------
\section{General properties of RRL transitions at ALMA frequencies}
\label{sec:rrl_properties}

Since the capabilities of ALMA for observing RRL transitions
have not been studied in detail yet, we make some general
remarks on the properties of recombination line observations with ALMA.
Once fully operational, ALMA will observe at ten bands with
wavelengths from $9.7\,$mm to $0.3\,$mm (frequencies of $31$ to
$950$\,GHz, see Table~\ref{alma}). Note that ALMA band~1 overlaps with
the upper frequency range of the EVLA.
The angular resolution in column~4 gives the expected range
in synthesized beams between the most compact and most extended array
configurations. The line sensitivity depends on the synthesised beam
size (angular resolution), as well as on the receiver sensitivity,
weather conditions (characterized by the precipitable water vapour
content), the integration time and the spectral resolution. The
values in column~6 correspond to the sensitivities assuming default
weather conditions for that frequency (determined from the ALMA
Observing Tool\footnote{http://almascience.eso.org/call-for-proposals/observing-tool/observing-tool}),
an integration time of $1\,$min and a spectral resolution of
$1$\,km\,s$^{-1}$. The two sensitivies correspond to values
for the compact and the most extended array configuration.
Table entries marked with ``$*$'' correspond to ALMA bands that are
still under development. The data for this table are taken from the ALMA Early
Science Primer\footnote{http://almatelescope.ca/ALMA-ESPrimer.pdf}.

%------------------------------
\begin{table*}
\caption{ALMA full array specifications\label{alma}}
\begin{tabular}{cccccc}
\hline
\hline
band & wavelength & frequency               & angular resolution & continuum sensitivity   & line sensitivity\\
     & (mm)       & (GHz)                   & (arcsec)           & (mJy/beam)              & (K)\\
\hline
 1 & $6.7$--$9.5$ &  $31.3$--$45$           & $13$--$0.1$        & $*$     &  $*$/$*$\\
 2 & $3.3$--$4.5$ &  $67$--$90$             & $6$--$0.05$        & $*$     &  $*$/$*$\\
 3 & $2.6$--$3.6$ & $84$--$116$             & $4.9$--$0.038$     & $0.05$  & $0.07$/$482$ \\
 4 & $1.8$--$2.4$ & $125$--$163$            & $3.3$--$0.027$     & $0.06$  & $0.071$/$495$ \\
 5 & $1.4$--$1.8$ & $163$--$211$            & $*$                & $*$     & $*$/$*$\\
 6 & $1.1$--$1.4$ & $211$--$275$            & $2.0$--$0.016$     & $0.10$  & $0.104$/$709$ \\
 7 & $0.8$--$1.1$ & $275$--$373$            & $1.5$--$0.012$     & $0.20$  & $0.29$/$1128$ \\
 8 & $0.6$--$0.8$ & $385$--$500$            & $1.07$--$0.009$    & $0.40$  & $0.234$/$1569$\\
 9 & $0.4$--$0.5$ & $602$--$720$            & $0.68$--$0.006$    & $0.64$  & $0.641$/$4305$\\ 
10 & $0.3$--$0.4$ & $787$--$950$            & $0.52$--$0.005$    & $1.2$   & $0.940$/$*$\\
\hline
\end{tabular}
\medskip\\
Table entries marked with ``$*$'' are not yet available.
\end{table*}
%------------------------------

To systematically investigate the properties of recombination lines at
ALMA frequencies, we have generated a list of all
recombination lines with $20 \leq n \leq 200$ and $1 \leq \Delta n
\leq 6$. Neglecting stimulated emission, the line strength will be
determined by the line emissivity (see equation~\eqref{eq:jldef} in Section~\ref{sec:physrecomb}).
Hence, we use the quantity $S_\mathrm{L} = N_m A_{m,n} =
N_m b_m N_{m}^{\mathrm{LTE}}$ with $m = n + \Delta n$ as a measure of
the line strength. We assume an electron number density $\nel = 5
\times 10^4\,$cm$^{-3}$ and a temperature $T = 10^4\,$K inside the
\hii\ region. For these fiducial values, we calculate $S_\mathrm{L}$
for the recombination lines, focusing on the relatively unexplored
ALMA frequency range.

\begin{figure}
\centerline{\includegraphics[height=170pt]{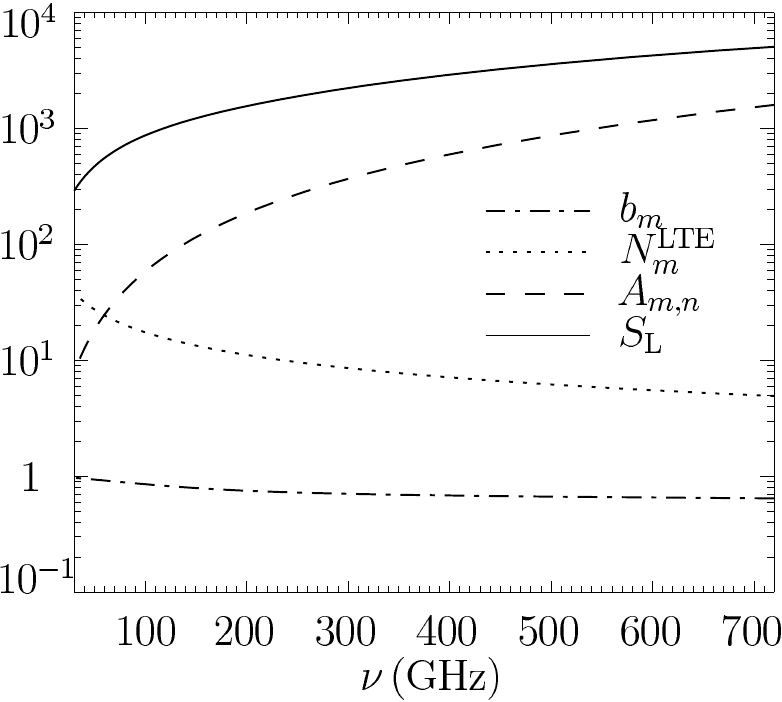}}
\caption{The departure coefficient $b_m$, the level population
  $N_{m}^{\mathrm{LTE}}$ (scaled by a factor of $10^{10}$), Einstein
  coefficient $A_{m,n}$ and line strength $S_\mathrm{L}$ (scaled by a
  factor of $10^{10}$) for $\alpha$-transitions in an \hii\ region
  with $\nel = 5 \times 10^4\,$cm$^{-3}$ and $T = 10^4\,$K for
  frequencies between 30 and 720 GHz.}
\label{fig:slcontr}
\end{figure}

It is interesting to note that two effects
partially cancel each other in their contribution to the line
strength. On the one hand, the Einstein coefficient $A_{m,n}$ grows
with frequency because larger frequencies at fixed $\Delta n$
correspond to smaller quantum numbers $m$. On the other hand, the
states with smaller quantum numbers are less occupied under
\hii\ region conditions than states with higher quantum numbers, so
the occupation number $N_{m}^{\mathrm{LTE}}$ falls off with increasing
frequency. The simultaneous consideration of these two effects
(together with a small decrease of the departure coefficient $b_m$
with growing frequency) leads to an increase in the line strength
$S_\mathrm{L}$ of more than one order of magnitude from the lowest to
the highest frequency ALMA bands. As an illustration,
Figure~\ref{fig:slcontr} shows the contributions of all three factors
to $S_\mathrm{L}$ across the ALMA frequency coverage.

As Figure~\ref{fig:slcontr} indicates, recombination lines in the
higher frequency bands will be stronger than for lower
frequencies. However, as shown in Table~\ref{alma}, the expected ALMA
sensitivity will decrease with increasing frequency. Since both the
line strength and the expected thermal noise grow roughly by a factor
of ten across the ALMA frequency range, we expect the recombination
lines of a fixed transition order $\Delta n$ in all ALMA bands to be equally
detectable.

\begin{figure*}
\includegraphics[width=500pt]{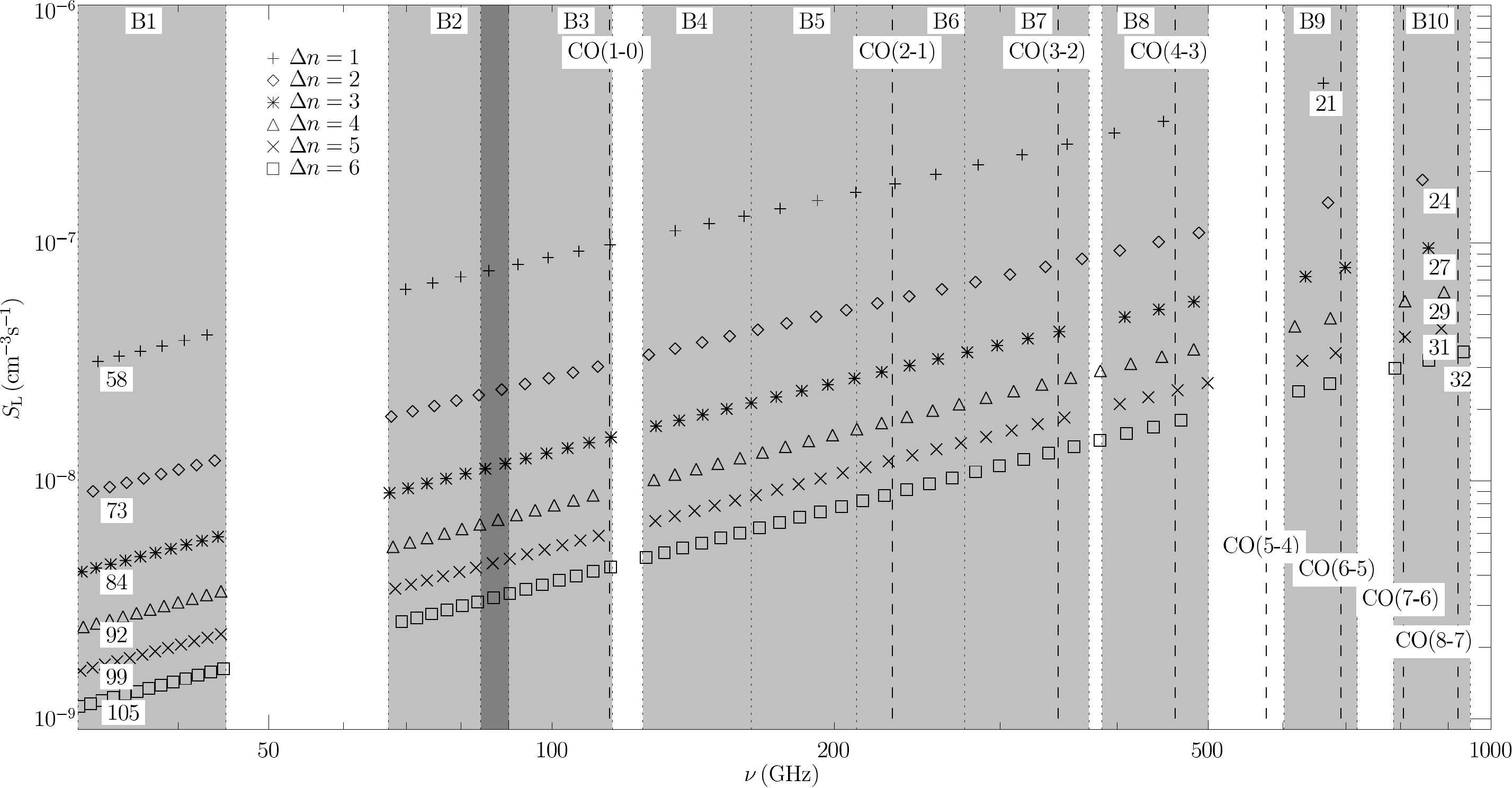}
\caption{Intensity of hydrogen recombination line transitions in the
  ALMA bands. The shaded regions mark the (partly overlapping) ALMA
  bands. The dashed lines show the location of some CO
  transitions. The numbers on the left- and right-hand side indicate
  the lower quantum number $n$ for the first and last transition for a
  series of fixed $\Delta n$, respectively.}
\label{fig:transplot}
\end{figure*}

The rest frequencies of individual transitions in the ALMA bands are
plotted in Figure~\ref{fig:transplot}.  We have also marked the rest
frequencies of some important transitions of the CO molecule. For
orientation, we show the lower principal quantum number $n$ for the
first and last transitions in a given transition order (at the left
and right edges of the plot). As expected, the lines become
brighter as the transition order $\Delta n$ decreases. The lowest
intensity of a line with transition order $\Delta n$ is (almost)
always larger than the highest intensity line with transition order
$\Delta n + 1$ in any single ALMA band, but not necessarily if several
bands are combined. As the figure shows, there are many transitions
covering the entire ALMA frequency range, so there will likely be a
recombination line in any wideband frequency setup. The recombination
lines are distributed almost equally with the logarithm of
frequency. For the convenience of the reader, we provide tables
containing all lines shown in Figure~\ref{fig:transplot} as
supplementary online material.

%------------------------------
%------------------------------

We note that the true line brightness can only be determined by
radiative transfer calculations through model \hii\ regions that also
take stimulated emission and absorption into account. We present such
calculations in Section~\ref{sub:sim_obs}. For one example \hii\ region,
\citet{gorsor02} calculate the effect of stimulated emission under non-LTE
conditions and find line depression for $\alpha$-transitions in the ALMA
range, with higher-frequency lines being more depressed than lower-frequency
lines. However, this effect is not strong enough to interfere with the
general trend shown in Figure~\ref{fig:transplot}.

%------------------------------------------------------------
\section{Asymmetric line profiles in mm and sub-mm RRL spectra}
\label{sec:asymm_profiles}

One interesting feature of hydrogen recombination lines in the
millimeter and sub-millimeter wavelength regime, as opposed to
recombination lines at centimeter wavelengths, is the prevalence of
non-LTE conditions. Appendix~\ref{sec:symm} explains how this can
give rise to asymmetric line profile shapes. Essentially, the non-LTE
effect is similar to a temperature gradient, and the resulting line
profiles resemble the well-known asymmetric (P-Cyg or inverse P-Cyg)
profiles which result from optically thick lines in expanding or collapsing
envelopes with a temperature gradient. As the non-LTE emission
becomes optically thick, the shape of the asymmetries can provide
important information about systematic motions in the gas. Therefore,
it is useful to know which physical conditions will cause a given line
to i) deviate from LTE and ii) become optically thick.

\begin{figure*}
\includegraphics[width=500pt]{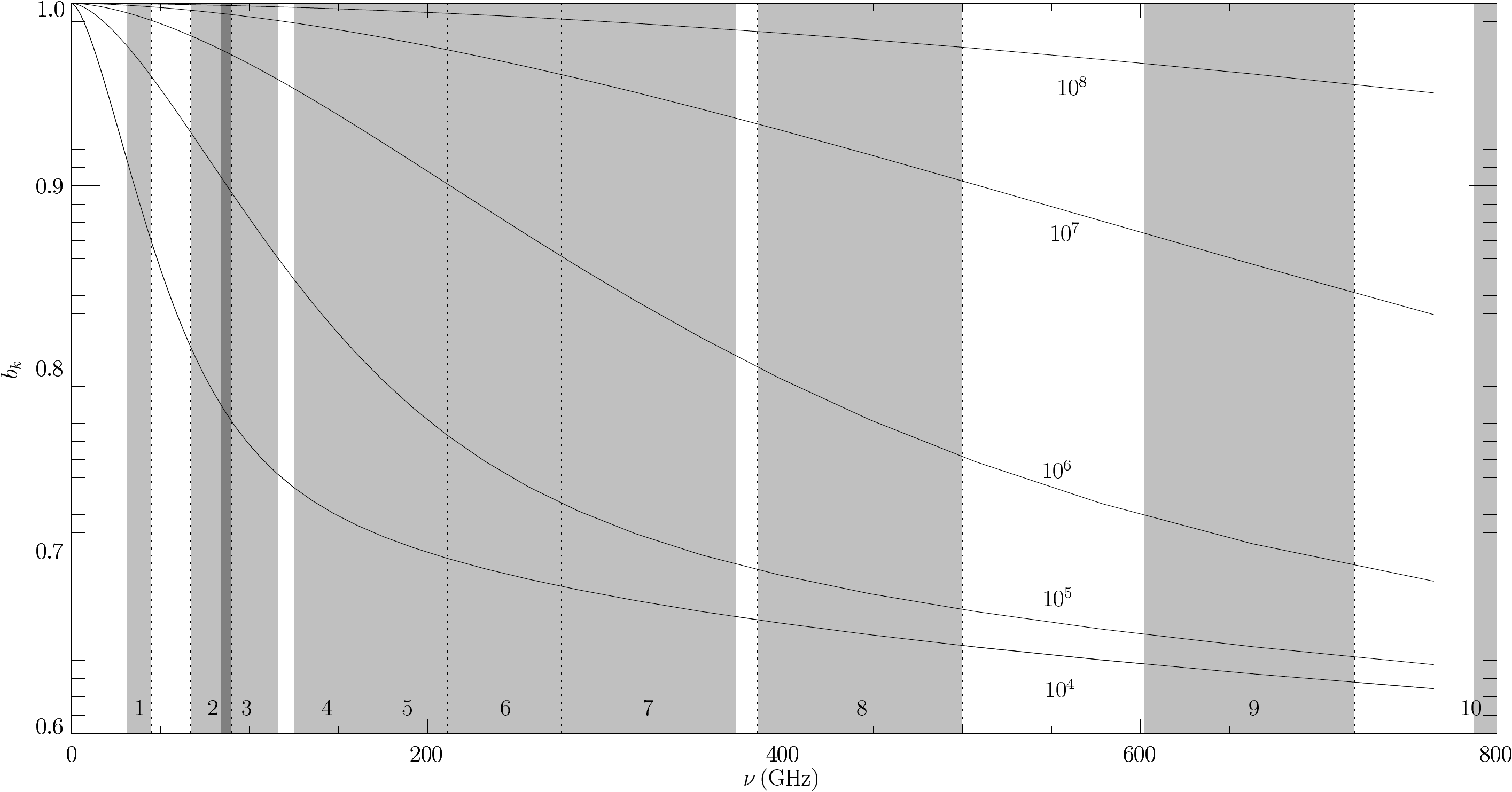}
\caption{Departure coefficients $b_k$ as function of frequency for
  transitions with $\Delta n = 1$ and electron number densities
  between $10^4$ and $10^8$\,cm$^{-3}$. The data is the same as in
  Figure~\ref{fig:walms1}. The ten ALMA bands are marked as vertical
  bands (as shown in Figure~\ref{fig:transplot}, there is no
  $\alpha$-transition in band 10 for the selected quantum
  numbers). The plot shows that departure from LTE conditions becomes
  increasingly relevant at the higher ALMA frequencies.}
\label{fig:almabk}
\end{figure*}

The deviation of a line from LTE is determined by the departure
coefficient $b_k$ (see Equation~\eqref{eq:bk} in Section~\ref{sec:physrecomb}).
Values further from one correspond to larger
deviations from LTE. Figure~\ref{fig:almabk} shows the departure coefficients
$b_k$ as function of frequency for transitions with $\Delta n = 1$
(giving a unique relationship between $k$ and $b_k$) and electron
number densities between $10^4$ and $10^8$\,cm$^{-3}$. The shaded
regions mark the ALMA bands. It becomes clear that departures from LTE
at typical \hii\ region densities \citep{kurtz05} for hypercompact
($\nel \gtrsim 10^6\,$cm$^{-3}$), ultracompact ($\nel \gtrsim
10^4\,$cm$^{-3}$) and compact ($\nel \gtrsim 10^3\,$cm$^{-3}$) regions
become larger at frequencies above $100\,$GHz, or wavelengths
shorter than $0.3$\,cm.

Departure coefficients smaller than unity are a necessary, but not sufficient
condition to see strong non-LTE effects. The emission optical depth
plays an important role too, as optically thin emission will not give
rise to line profile asymmetries. Determining whether strong
asymmetries are produced requires radiative transfer modelling of the
RRL emission. In Section~\ref{sub:sim_obs} we generate simple analytical
\hii\ regions, solve the non-LTE radiative transfer equations for RRLs
at frequencies covered by the EVLA and ALMA, then simulate observing
these regions for typical EVLA and ALMA observing setups.

%------------------------------
%------------------------------

%------------------------------------------------------------
\subsection{Simulated EVLA and ALMA observations}
\label{sub:sim_obs}

We aimed to simulate RRL observations of \hii\ regions, covering a
range of typical EVLA and ALMA observational setups. Synthetic
spherically symmetric \hii\ regions were generated
using analytical profiles for $\nel$ (equation~\eqref{eq:model1}) and the radial velocity $v_r$ (equation~\eqref{eq:model6}),
corresponding to model 6 from Appendix~\ref{sec:tests}, which were chosen as representative of compact \hii\
regions. We then used the radiative transfer modelling to generate
synthetic emission maps of the \hii\ regions.  The radiative transfer
models deal with continuum and line emission simultaneously so create
a combined line plus continuum emission cube. We separated the two
components by fitting a low-order polynomial to the line-free channels
and from there on treated each component individually. In this way
synthetic continuum and RRL emission cubes were generated for
transitions at frequencies which lie in each of the standard EVLA and
ALMA cm to sub-mm observing windows. The angular scale and flux were
corrected for a source distance of 3kpc---a typical distance to
nearby \hii\ regions.

\begin{figure}
\includegraphics[height=200pt]{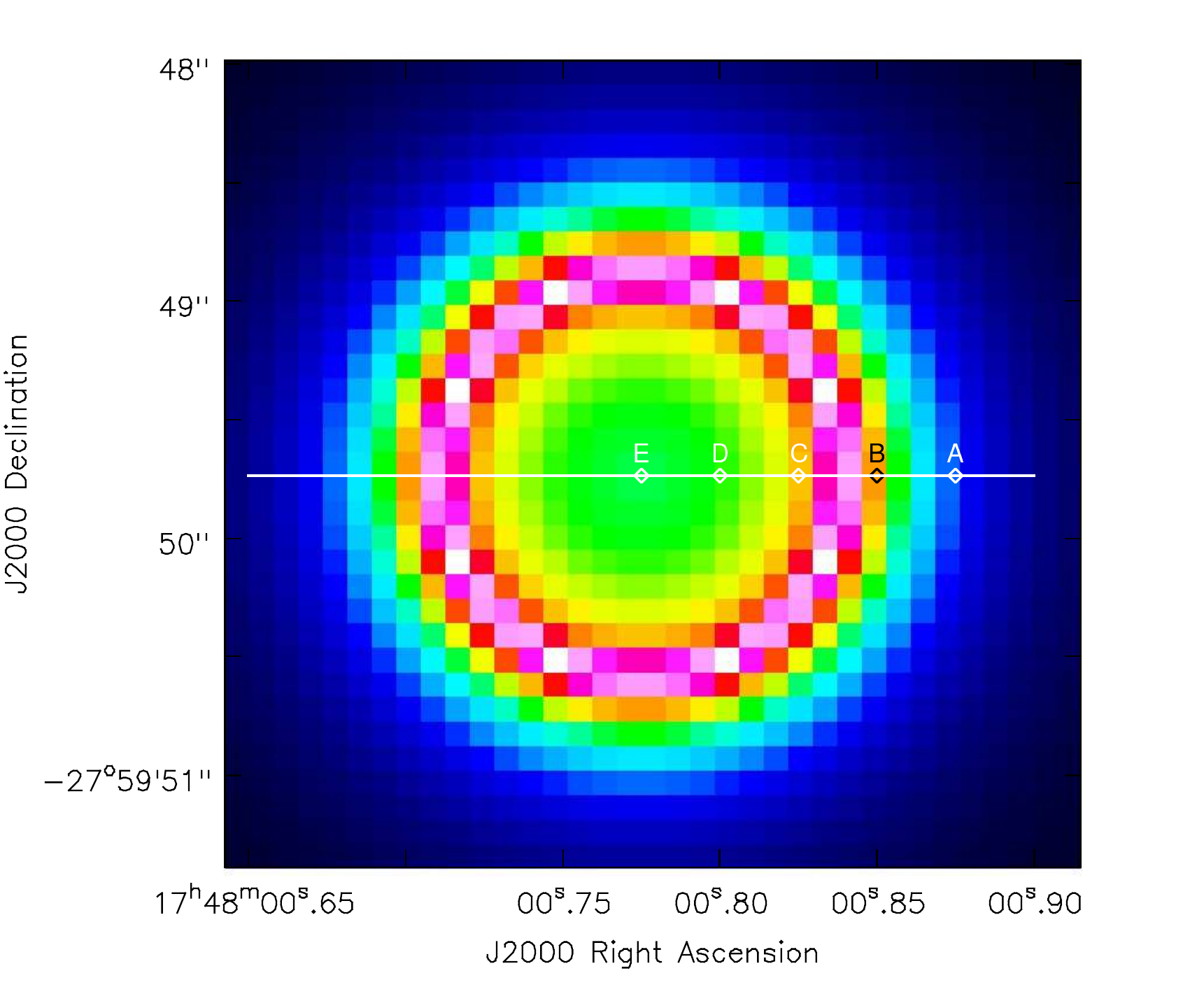}
\caption{Integrated intensity map of the synthetic H68$\alpha$
   emission line cube (see Section~\ref{sub:sim_obs} for details). The horizontal
   white line shows the cut used to create the position-velocity
   diagrams in Figure~\ref{fig:pv}. The points A to E are positions at which
   spectra are extracted for Figures~\ref{fig:spectra} and~\ref{fig:atoe}. Point E indicates the
   centre of the \hii\ region. The colour scale shows the integrated
   intensity of the line emission initially increases towards the
   centre. However, between points B and C the emission becomes
   optically thick and the integrated intensity drops with decreasing
   radius.}
\label{fig:h68a}
\end{figure}

The synthetic line emission cubes at all frequencies have a similar
morphology. Figure~\ref{fig:h68a} shows the integrated intensity map
of the synthetic H68$\alpha$ line emission cube as an example.
At radii larger than a few arcseconds the RRL emission is
optically thin and the profiles are symmetric. At smaller radii
(approaching $\sim$1$\arcsec$) the emission intensity increases but
starts to become optically thick so the profiles begin to deviate from
symmetric. At a radius of $\sim$1$\arcsec$ the emission becomes
optically thick. The intensity then drops sharply and the profile
asymmetries become more pronounced with decreasing radius. The
transition from optically thin to thick emission is seen as a very
pronounced boundary in the RRL integrated intensity maps.

%------------------------------
%------------------------------

We then simulated observing the synthetic maps with the EVLA and
ALMA. For each telescope we chose a set of observational parameters
(integration time, array configuration, correlator setting etc.) that
represent the expected range in these parameters for typical
observations. We chose integration times of 0.5 hours and 8 hours to
represent snapshot imaging and deep integrations, respectively. We
selected array configurations which gave synthesized beams close to
either 1$\arcsec$, 0.3$\arcsec$ or 0.1$\arcsec$ (corresponding to
physical scales of 15, 4.4 and 1.5\,mpc, respectively), including both
compact and extended array configurations to investigate the effect of
spatial filtering and surface brightness sensitivity. 

For each observational setup we then used the CASA task SIMDATA to
generate synthetic visibilities (measurement sets) of the sky
model. For the ALMA observations, thermal noise was added using the
``tsys-atm'' parameter, which models the typical atmospheric
conditions above the ALMA site for a given precipitable water vapour
(PWV) content. The PWV for each observing frequency was selected to be
represenative of the conditions expected for observations carried out
at that frequency (see Table~\ref{sim_obs_pars}). At present the
``tsys-atm'' mode only models the atmosphere above the ALMA site. For
the EVLA observations we therefore used the ``tsys-manual'' mode
taking the typical zenith opacities as a function of frequency from
the AIPS\footnote{http://www.aips.nrao.edu/index.shtml} task CLCOR,
assuming a ground temperature of 11.7\,C (the yearly median surface
temperature at the VLA site) and a sky temperature of
275\,K\footnote{These values were taken from the VLA test memo number 232:
http://www.vla.nrao.edu/memos/test/232/232.pdf}. We note in
passing that corruption of phases using ``tsys-atm'' will be worse
than the errors in real ALMA data because the water vapour radiometers
should partially correct for this. We also note that the above
modelling does not take into account potential line confusion which
will be an issue for some lines, particularly at higher frequencies. A
sampling time of 10 seconds was used throughout for both EVLA and ALMA
observations.

The synthetic visibilities were inverted, cleaned and restored using
the CASA task CLEAN. Dirty images were created with an area large
enough to cover the model emission and with a pixel scale of $\sim$1/3
the synthesised beam. Mode ``MFS'' with index = 1 (assuming no
frequency dependence on the flux of the continuum emission) and
``channel'' were used to clean the continuum and line cubes,
respectively. A clean box was created based on the known structure of
the model emission and a first-pass clean was performed on the cubes
using a threshold of 3 times the theoretically predicted thermal
noise. These results were then inspected and, where necessary,
optimised by additional manual cleaning. All the images were corrected
for primary beam attenuation as this becomes important at the higher
observing frequencies where the primary beam is comparable to the
angular extent of the model emission.

%------------------------------
\begin{table}
  \caption{Telescope, radio recombination line transitions and
    atmospheric parameters used for the simulated
    observations. See text for details. \label{sim_obs_pars}}
\begin{tabular}{ccccc}\hline\hline

Telescope & Transition    &  Frequency    & PWV   & $\tau_\mathrm{zenith}$   \\
          &               &   (GHz)       & (mm)  &                   \\ \hline
EVLA      &  H186$\alpha$ & 1.013767      & -     & 0.008\\  
EVLA      &  H117$\alpha$ & 4.053878      & -     & 0.01\\
EVLA      &  H93$\alpha$  & 8.045603      & -     & 0.01 \\
EVLA      &  H68$\alpha$  & 20.46177      & -     & 0.05 \\
EVLA      &  H63$\beta$   & 50.19619      & -     & 0.05  \\
ALMA      &  H39$\alpha$  & 106.7374      & 2.7   & -  \\
ALMA      &  H30$\alpha$  & 231.9010      & 1.8   & -  \\
ALMA      &  H37$\gamma$  & 346.7585      & 1.3   & -  \\
ALMA      &  H26$\alpha$  & 353.6228      & 1.3   & -  \\
ALMA      &  H24$\alpha$  & 447.5403      & 1.3   & -  \\ \hline

\end{tabular}
\end{table}

%---------------
\subsection{Results}
\label{subsub:results}

We find the observational results depend critically on three
parameters: (i) whether the observations have sufficient angular
resolution to resolve the optically thick central region; (ii) whether
they have sufficient surface brightness sensitivity to detect this
optically thick line emission; (iii) whether the intrinsic line
profiles are narrow enough to spectrally resolve the asymmetries.

The issue with the intrinsic line profile being too wide to spectrally
resolve the line asymmetries dominates at lower frequencies because
the pressure broadening is such a steep function of frequency
(see Equation~\eqref{eq:delta} in Section~\ref{sec:physrecomb}).
This makes it impossible to detect the line
asymmetries in the H186$\alpha$ observations.

\begin{figure}
\includegraphics[height=200pt,angle=-90]{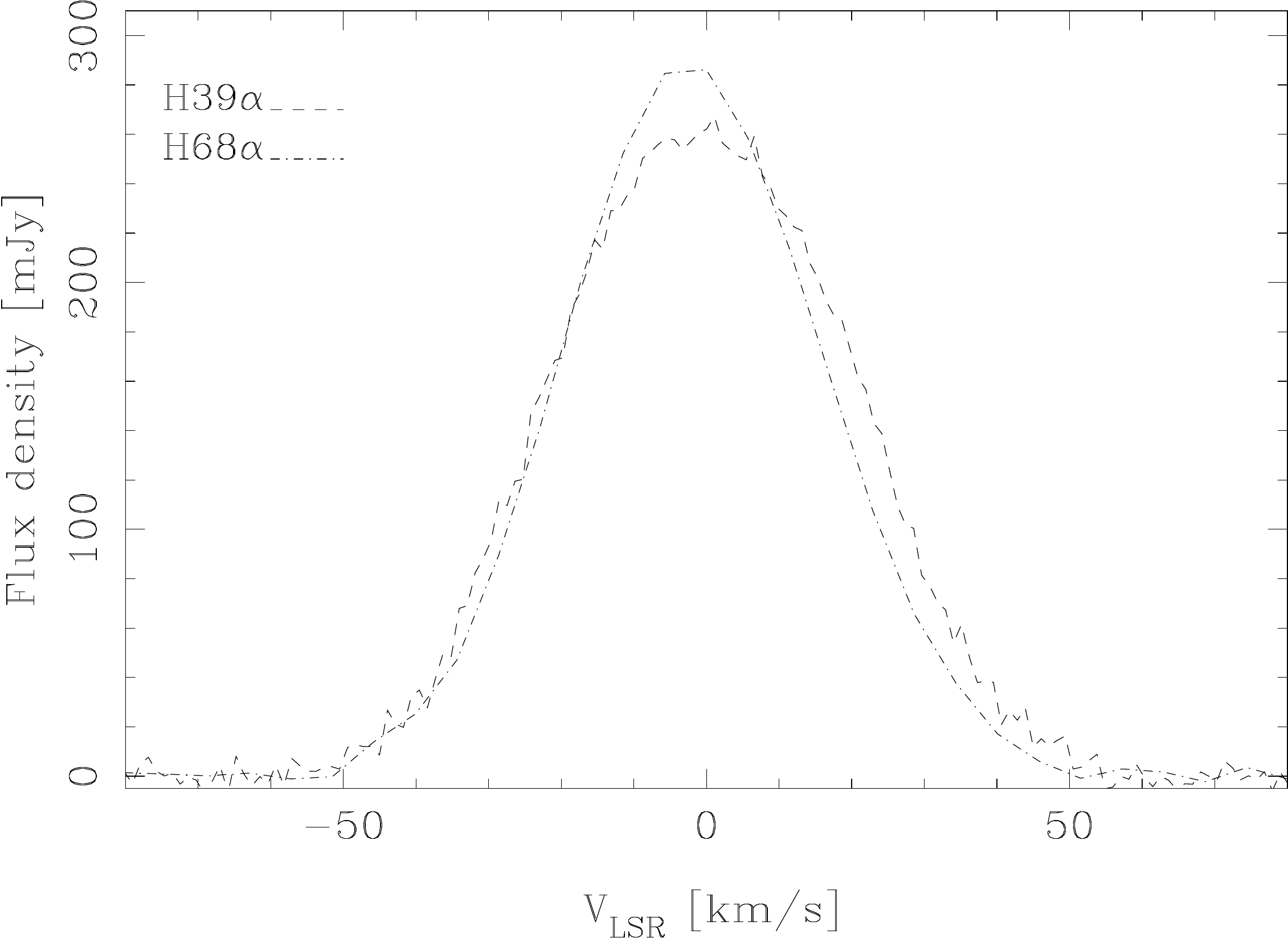}
\vspace{1.5cm}
\caption{H39$\alpha$ and H68$\alpha$ spectra extracted from the
    central position ("E" in Figure~\ref{fig:h68a}) of the 1$\arcsec$ angular resolution
    simulated observations (see Section~\ref{sub:sim_obs} for details). The line
    profiles for both transitions are nearly symmetric, even though
    the H68$\alpha$ line profile is clearly asymmetric in the model data
    cubes. This is because the 1$\arcsec$ synthesised beam is not sufficient
    to resolve the optically thick central region. However, the
    H68$\alpha$ line is shifted to lower velocities than the H39$\alpha$
    line, which peaks at the systemic velocity ($0\,$km/s). This is
    because the H68$\alpha$ emission is optically thick and, due to the
    systematic velocity structure, the red-shifted emission is
    preferentially self-absorbed, leading to a blue-ward shift in the
    peak of the emission. Searching for progressive velocity offsets
    from the systemic velocity with lower frequency (more
    optically thick) RRL transitions offers a way to infer systematic
    motions in the gas, even if it is not possible to resolve the
    optically thick region.}
\label{fig:spectra}
\end{figure}

As shown in Figure~\ref{fig:spectra}, the RRL emission is strongly detected in the 1$\arcsec$
observations. However, the line profiles for all transitions are
nearly symmetric, even though the lower frequency transitions are
clearly asymmetric in the model data cubes. This is because the
angular resolution is insufficient to spatially resolve the line
emission coming from the optically thin shell at large radii, from
the optically thick emission observed towards the center of the
source. As the optically thin emission is brighter and has a
larger solid angle, it dominates the optically thick emission when
they are both in the same synthesised beam. It is interesting to
note in Figure~\ref{fig:spectra} that the lower frequency H68$\alpha$ transition is
skewed by a few km/s to lower velocities than the higher frequency
H39$\alpha$ transition, which peaks at the systemic velocity
($0\,$km/s). The explanation for this becomes clear from the higher
resolution observations (see below for more details). The H68$\alpha$
emission is optically thick and, due to the systematic velocity
structure, the red-shifted emission is preferentially
self-absorbed, leading to a blue-ward shift in the peak of the
emission. Although the deviation from symmetric line profiles is
only slight, searching for progressive velocity offsets from the
systemic velocity with lower frequency (more optically thick) RRL
transitions offers a way to infer systematic motions in the gas,
even if it is not possible to resolve the optically thick
region. However, in order to clearly detect the line profile
asymmetries, observations should have sufficient angular
resolution to separate the optically thick and thin regions.
Such velocity offsets have already been found \citep[e.g.][]{ketoetal08}.

%------------------------------
%------------------------------

The 0.1$\arcsec$ observations suffer from a different problem. They
have sufficient angular resolution to resolve the optically thick
region but the surface brightness sensitivity in the 0.1$\arcsec$
synthesised beam is too low to detect the RRL emission, even for an 8
hour integration with the EVLA and ALMA.

\begin{figure}
\includegraphics[height=180pt,angle=-90]{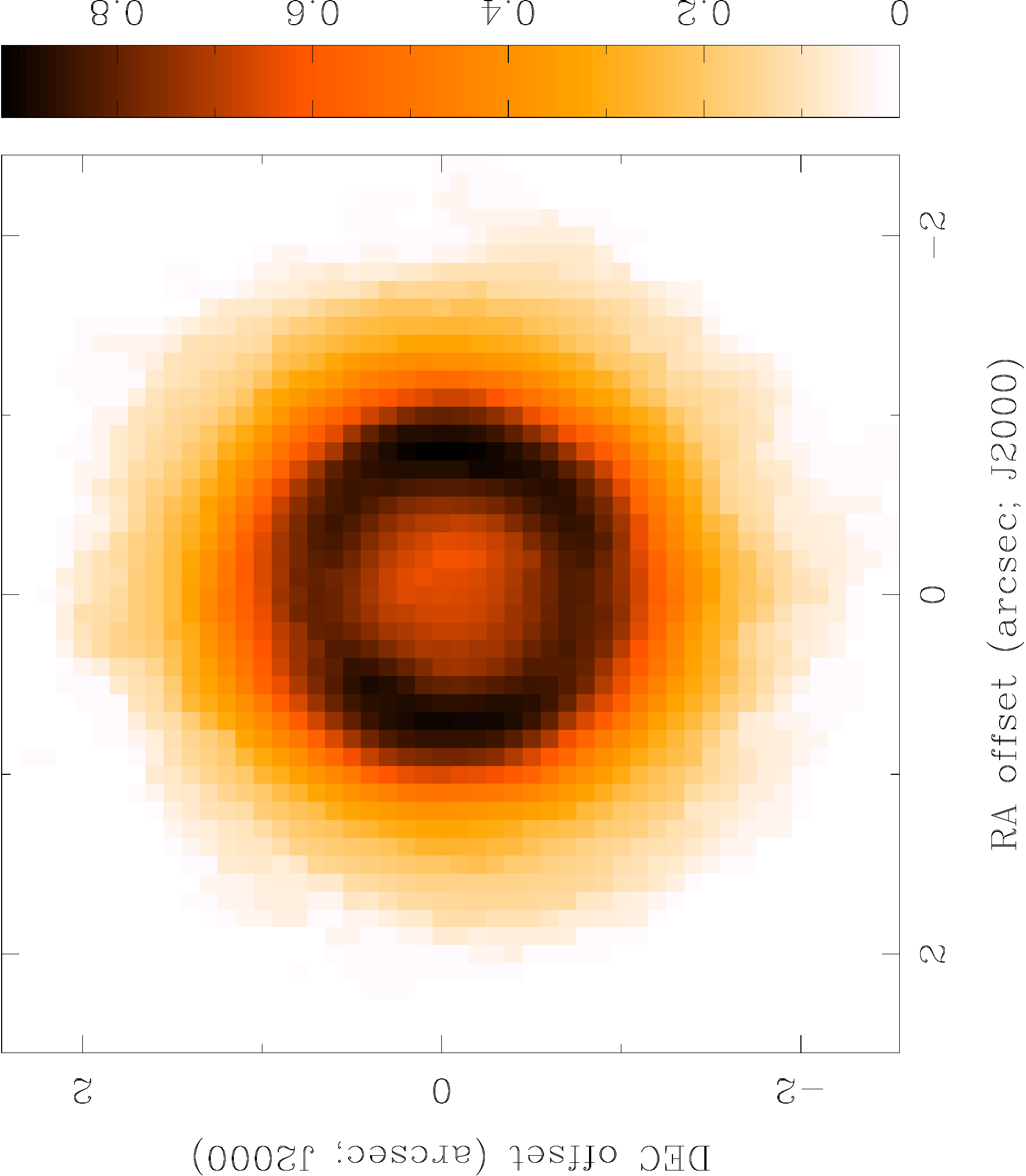}\\[17pt]
\includegraphics[height=180pt,angle=-90]{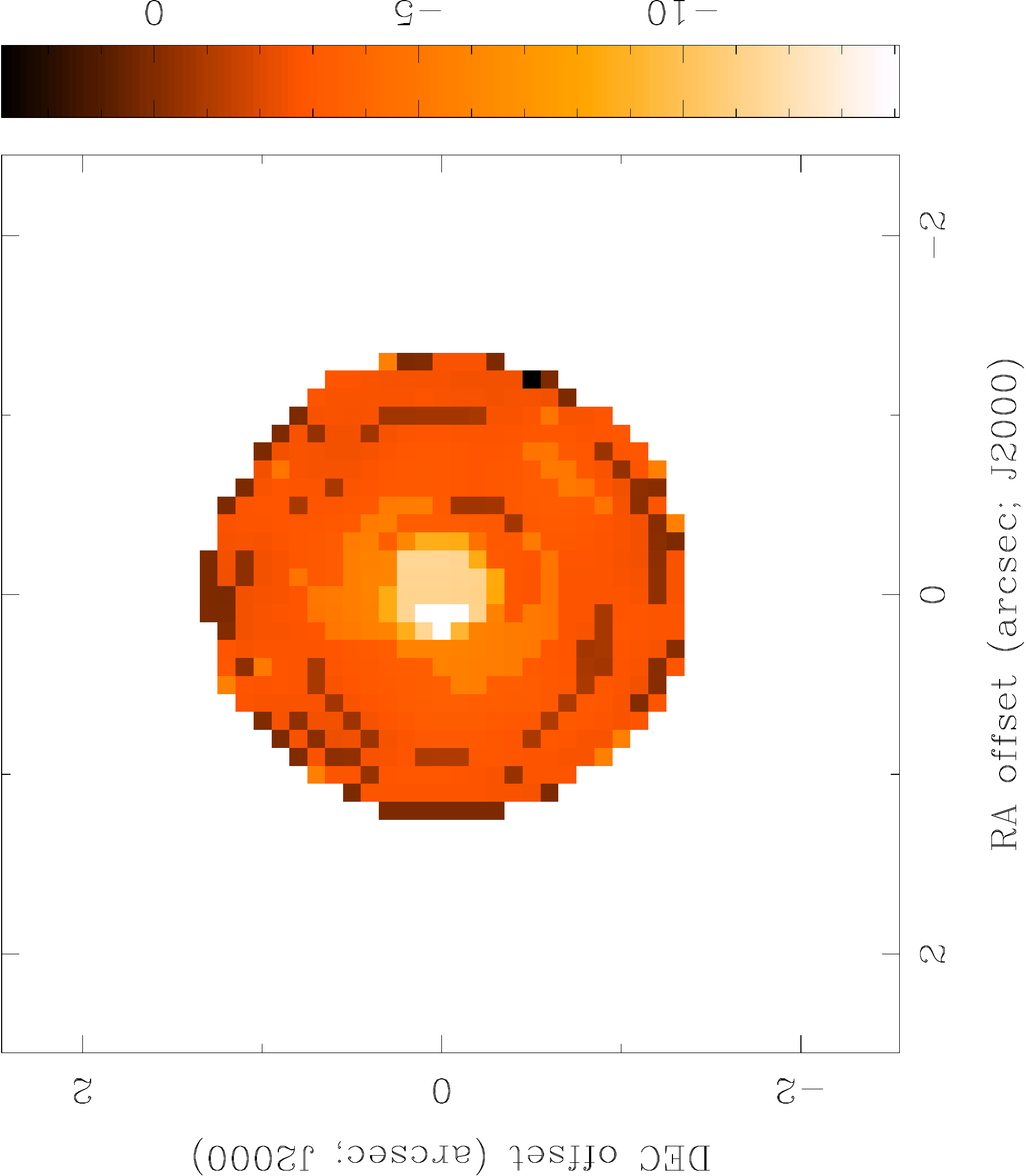}\\[17pt]
\includegraphics[height=180pt,angle=-90]{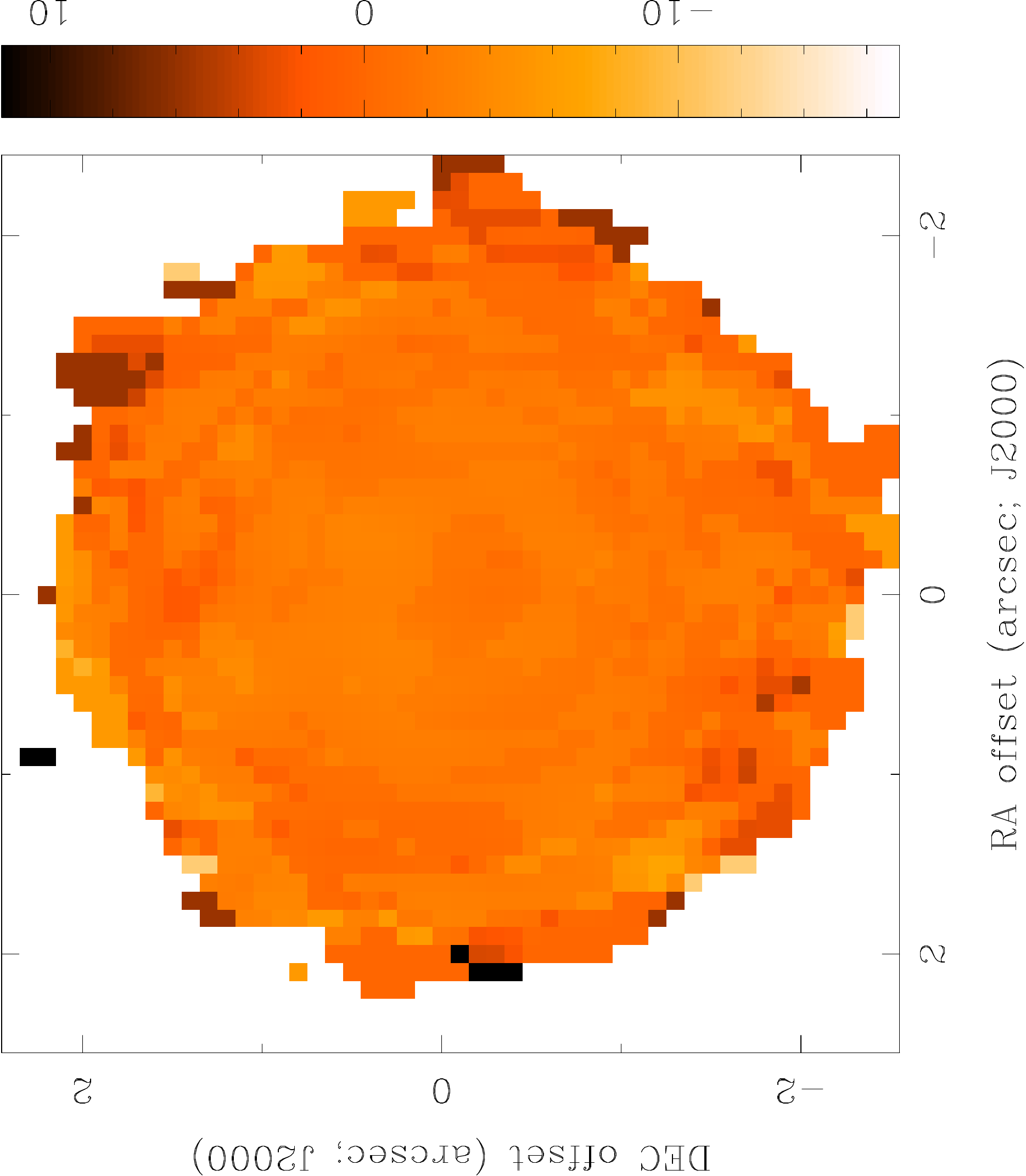}\\[5pt]
\caption{Integrated intensity (zeroth moment, top, with units Jy/beam$\times$km/s) and
    velocity-weighted intensity (first moment, center and bottom, with units km/s) maps of the simulated
    0.3$\arcsec$ H68$\alpha$ RRL observations. The two first moment maps have
    been made with the different intensity cuts. The central map
    was made using a threshold of
    10\,mJy (i.e. all pixels below 10\,mJy were excluded before the
    intensity-weighted velocity was calculated). This accentuates the
    brighter emission and a clear blue-shifted velocity offset is seen
    towards the center of the region, where the H68$\alpha$ emission
    becomes optically thick (see Section~\ref{sec:preditc_conditions} for more details). The
    bottom map was made with a
    much lower threshold of 3mJy. The lower intensity emission
    dominates the map. As a result, no velocity shift is seen.}
\label{fig:moment}
\end{figure}

The 0.3$\arcsec$ observations have both sufficient angular resolution
to resolve the optically thick region and sufficient surface
brightness sensitivity to detect the RRL emission.
Figure~\ref{fig:moment} shows  the integrated intensity (zeroth moment) and velocity-weighted
intensity (first moment) maps of the simulated 0.3$\arcsec$ H68$\alpha$ RRL
observations. The integrated intensity map shows these
observations have both sufficient surface brightness sensitivity
to detect the RRL emission and sufficient angular resolution to
resolve the optically thick region.

%------------------------------
%------------------------------

\begin{figure}
\includegraphics[height=160pt,angle=-90]{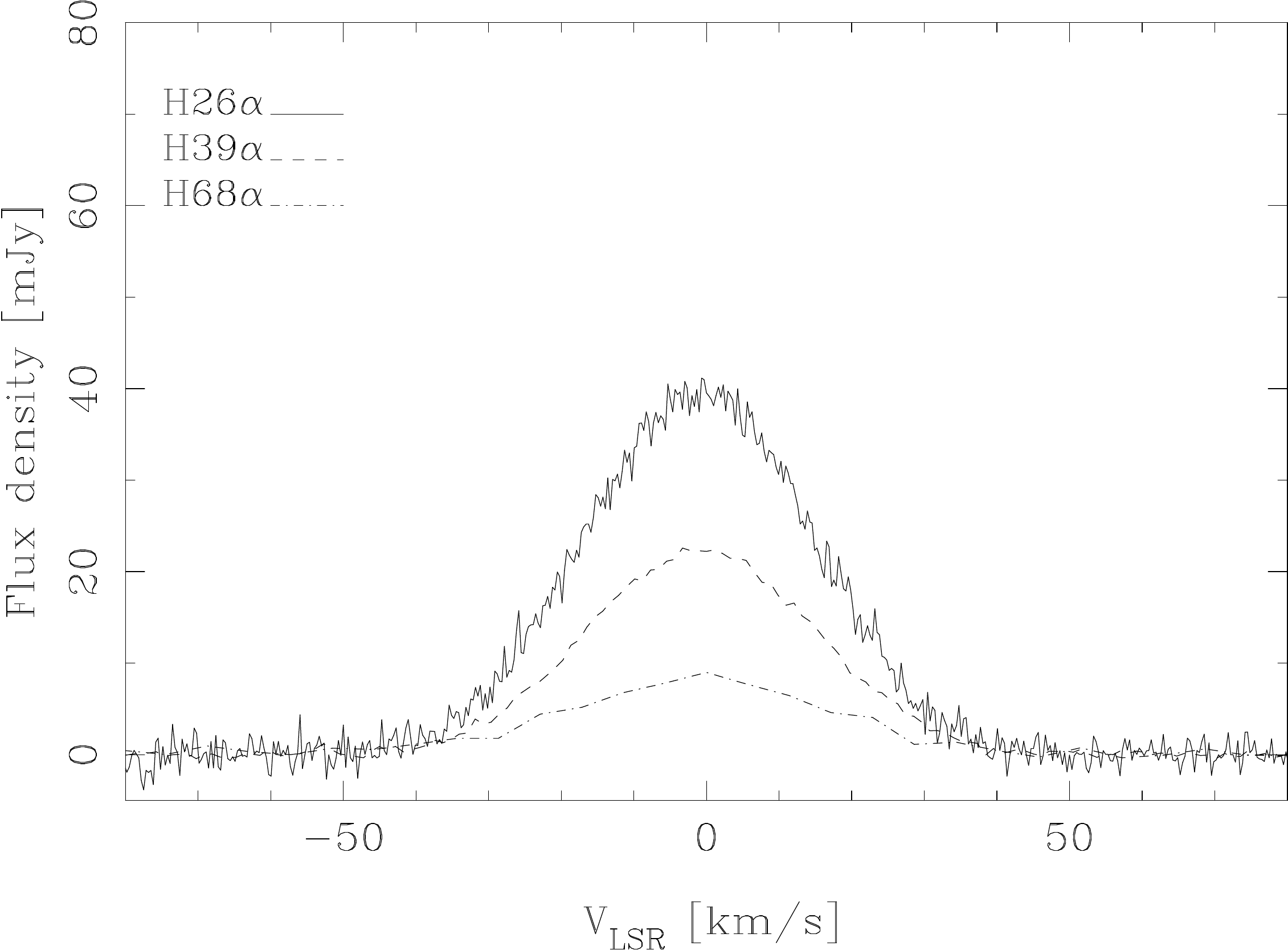}\\[15pt]
\includegraphics[height=160pt,angle=-90]{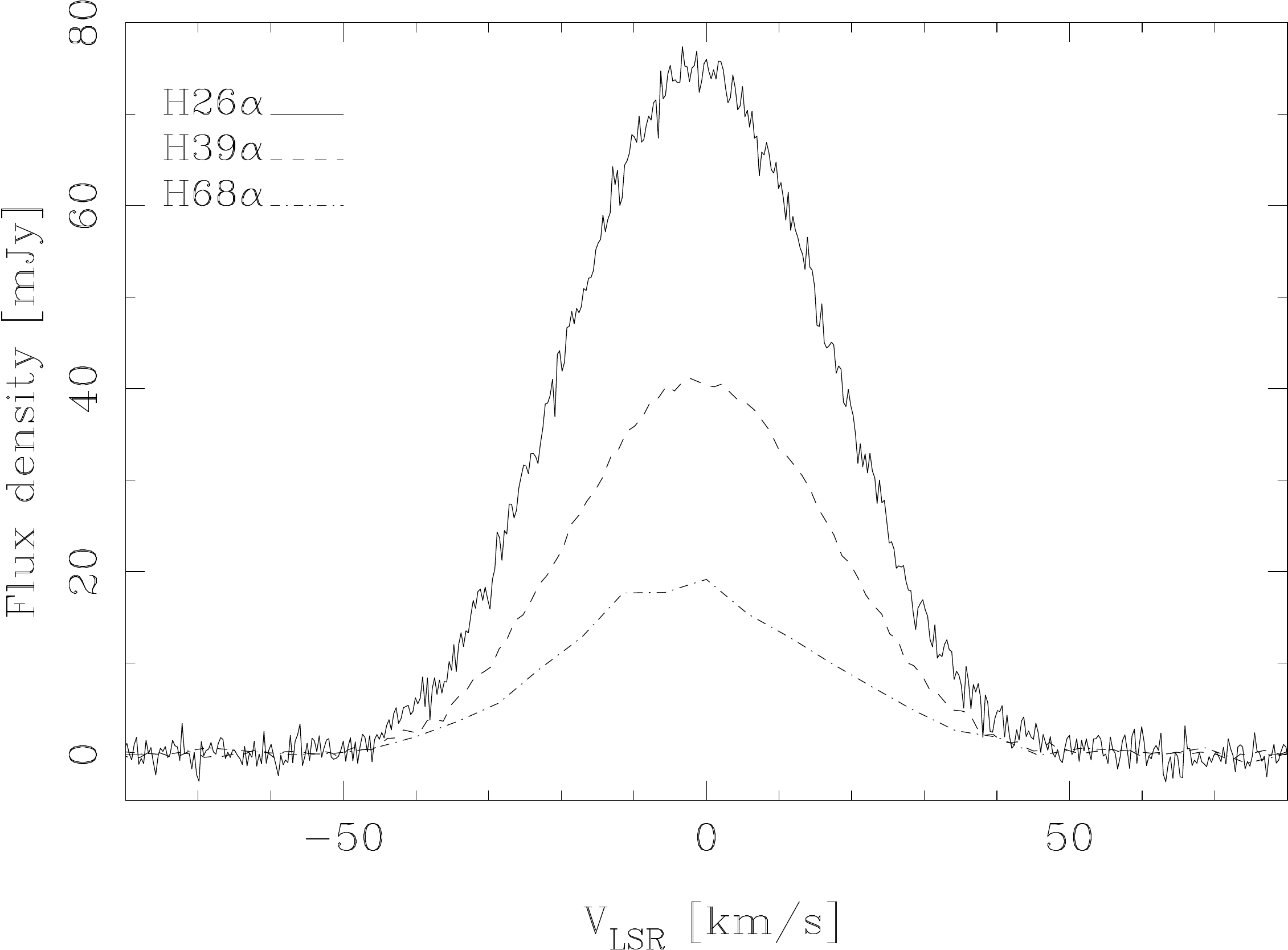}\\[15pt]
\includegraphics[height=160pt,angle=-90]{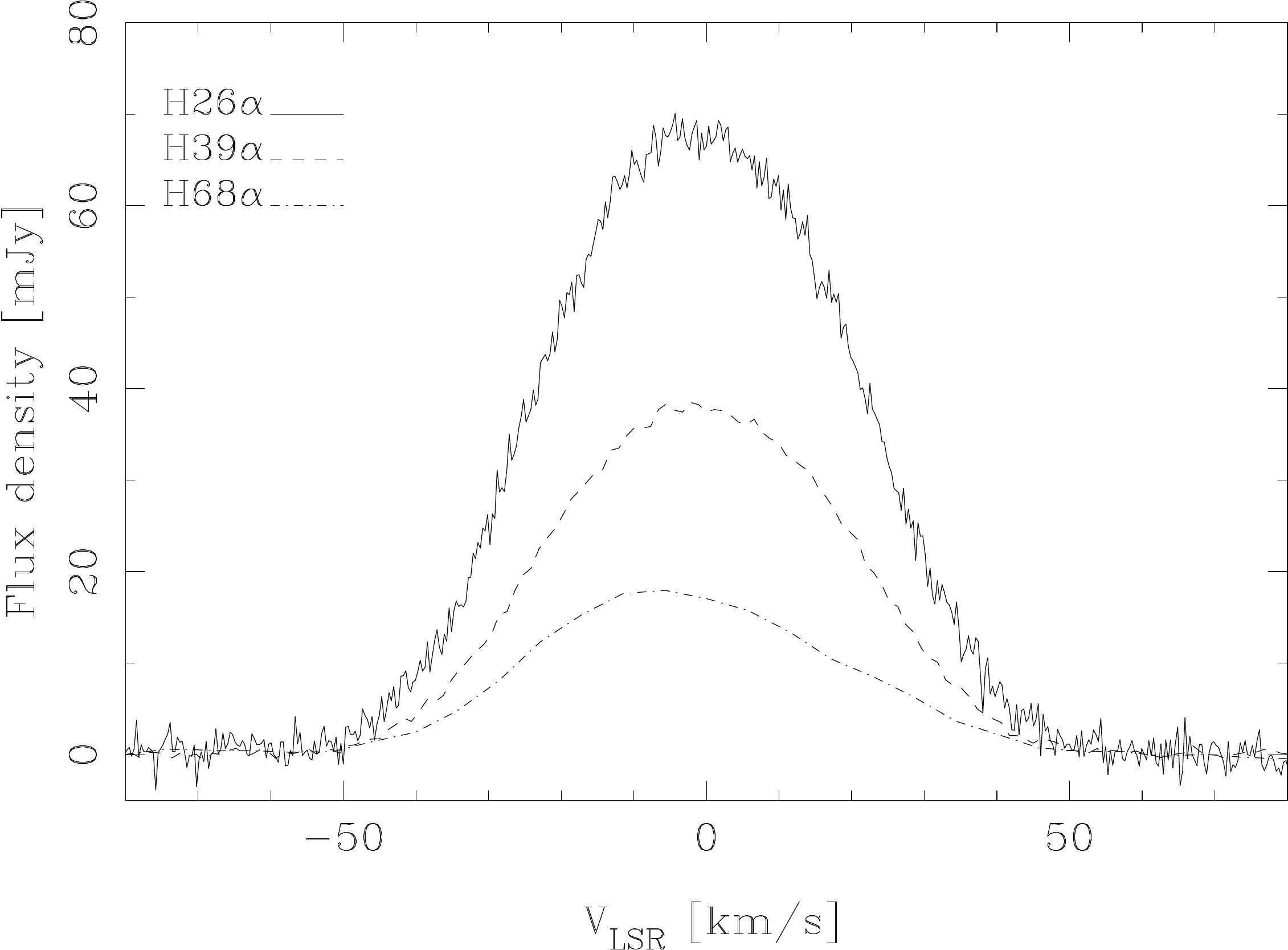}\\[15pt]
\includegraphics[height=160pt,angle=-90]{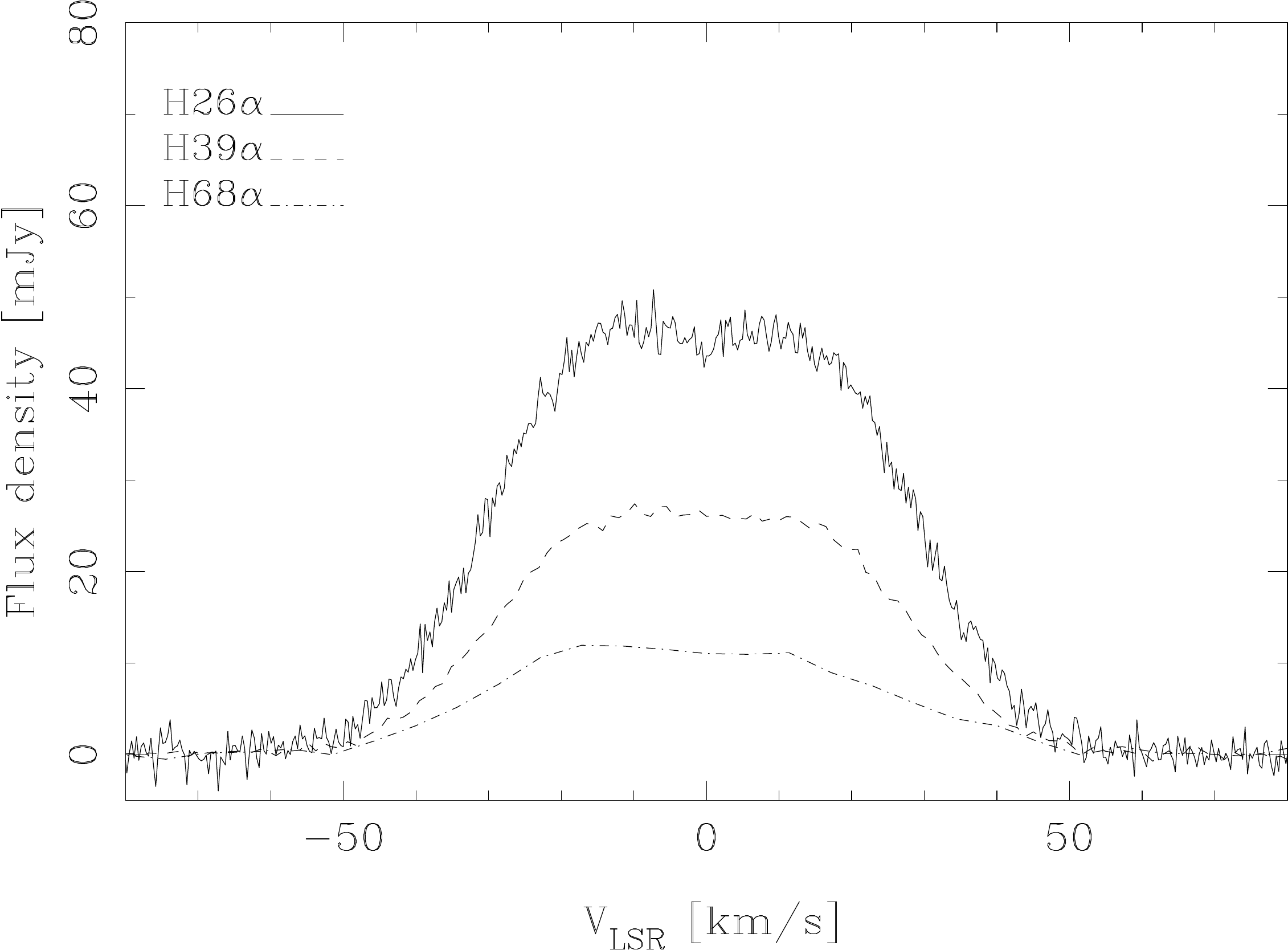}\\[15pt]
\includegraphics[height=160pt,angle=-90]{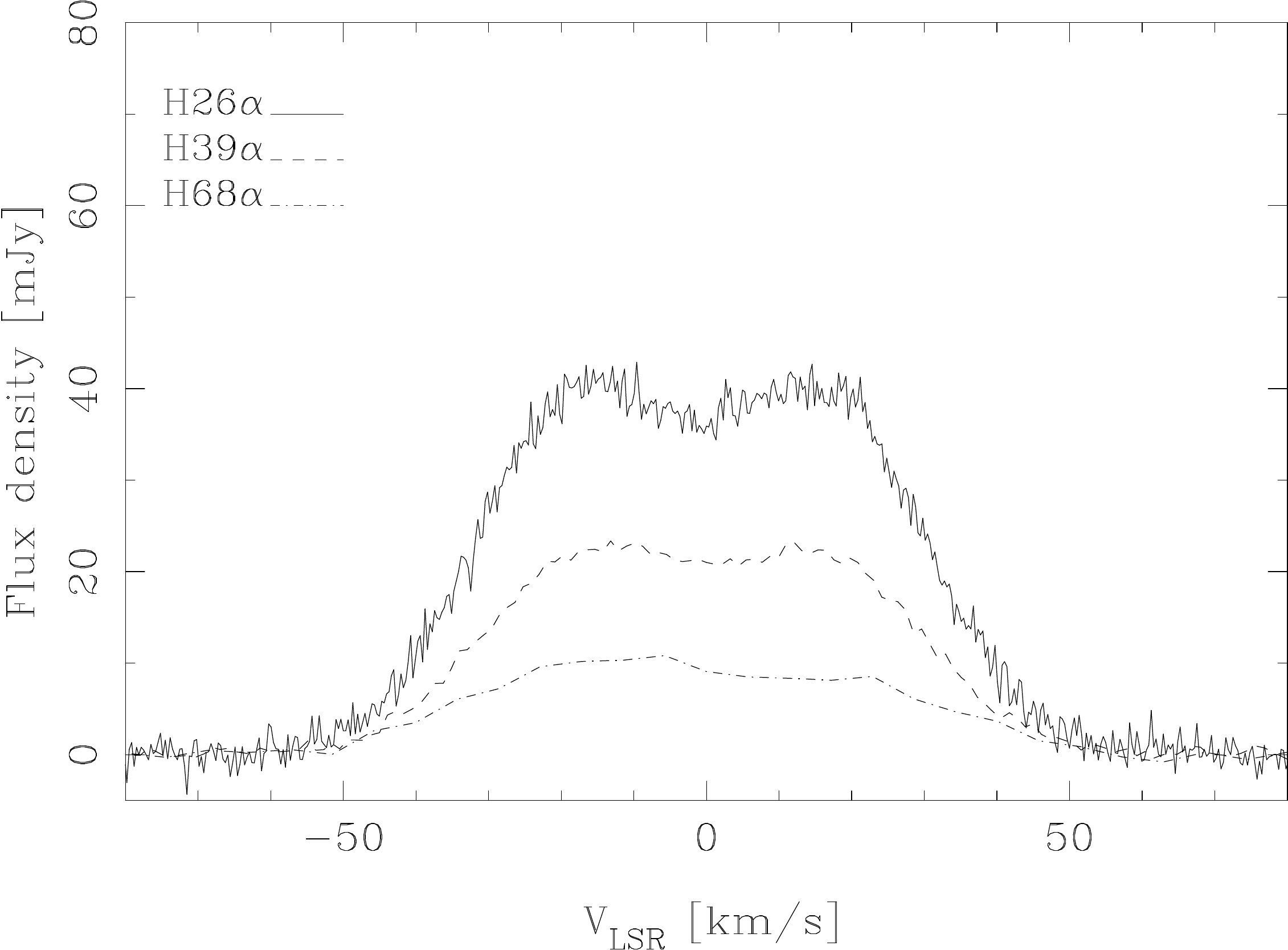}\\[15pt]
\caption{H68$\alpha$, H39$\alpha$ and H26$\alpha$ spectra extracted
    from positions A to E (top to bottom) in Figure~\ref{fig:h68a} for the 0.3$\arcsec$ resolution
    simulated observations.  All spectra are plotted on the same scale
    and the systemic velocity is 0\,km/s.}
\label{fig:atoe}
\end{figure}

Figure~\ref{fig:atoe} shows
the H68$\alpha$, H39$\alpha$ and H26$\alpha$ spectra extracted from
positions A to E in Figure~\ref{fig:h68a} for the 0.3$\arcsec$ resolution simulated
observations. All spectra are plotted on the same scale and the
systemic velocity is 0\,km/s. The highest frequency (H26$\alpha$)
transition is noticably brighter than the lower frequency
transitions. This offsets the higher thermal noise inherent in the
higher frequency observations. At the outer most position (A), the
emission is optically thin and the line profiles from all
transitions is symmetric and peaks at the same velocity. At
position B the flux density from all transitions increases. The
H39$\alpha$ and H26$\alpha$ spectra are still optically thin, so the
line profiles are symmetric and peak at 0\,km/s. However, the
H68$\alpha$ emission is becoming optically thick and a red-shifted
absorption shoulder is observed in the line profile. This effect
is more pronounced at position C, and the H68$\alpha$ spectra peak
velocity is blue-shifted to $\sim-10\,$km/s. The flux of the higher
frequency transitions drops compared to position B, showing they
are also being affected by increasing optical depth. At position D
the line profiles are all strongly self-absorbed. The measured
velocity dispersion of the lines at this position, for example as
measured by the full width at half maximum (FWHM), would be much
larger than at positions farther from the center. The self
absorption in all the spectra at the central position (E) is
strong enough to produce line profiles with two peaks separated by
tens of km/s. Without the line profiles from the optically thin
emission at larger radii it would not be possible to tell if this
emission was from a single, optically thick velocity component or
two seperate, optically thin velocity components.

%------------------------------
%------------------------------

It is important to note that the line profiles vary in a systematic
way as the line frequency changes. At a fixed position, the line profile
becomes increasingly asymmetric as the frequency decreases because the
optical depth increases, enhacing the non-LTE effects. This overcompensates
the larger value of the departure coefficients, which are getting closer
to the LTE-value of unity towards smaller frequencies. This change of
the line profiles with frequency is characteristic for the non-LTE
effect that can be used to trace systematic gas motions. 
Asymmetries\footnote{In principle,
asymmetries could also be produced by temperature gradients under
LTE conditions. However, detailed simulations of the formation and expansion
of ultracompact \hii\ regions that self-consistently calculate the temperature
in the ionized gas \citep{petersetal10a,petersetal10b,petersetal10c,petersetal11a}
show that the temperature is very homogeneously equal to $10^4\,$K across the \hii\ region.}
in the line profile due to asymmetries in the \hii\ region, however, are
expected to be equally pronounced at all frequencies. In reality, of
course, both effects will occur simultaneously. Therefore, detailed
radiative transfer modeling might be necessary to extract the desired
information on systematic gas motion like infall or outflow\footnote{In
this paper, the word outflow refers to the pressure-driven expansion
of the \hii\ region, not to bipolar outflows such as those driven from accretion
processes of young stellar objects.} inside the \hii\ region. Our
implementation of recombination lines in RADMC-3D is an ideal
tool to carry out such modeling.

The information
in the spectra of Figure~\ref{fig:atoe} can also be visualised in other
ways. The H68$\alpha$ velocity-weighted intensity (first moment) map
in the central panel of Figure~\ref{fig:moment}
shows a clear break in the
velocity structure. At large radii, where the emission is
optically thin, the intensity-weighted velocity is close to 0\,km/s---the
systemic velocity of the \hii\ region. Once the emission
becomes optically thick, the red-shifted emission is
preferentially absorbed due to the systematic velocity structure
(infall) of the gas. The fact there is more blue-shifted than
red-shifted emission means the velocity-weighted intensity
therefore shifts to $\sim-10\,$km/s. This characteristic 'bulls eye'
morphology is a typical infall/outflow signature in
intensity-weighted velocity maps. However, if the threshold used
to define the integrated intensity map is dropped from 10\,mJy to
3\,mJy (a few times the noise level: bottom panel of Figure~\ref{fig:atoe}),
the intensity-weighted
velocity is dominated by the optically thin, lower intensity
emission at higher velocities so the 'bulls eye' pattern
dissapears and a uniform velocity field at the systemic velocity
is seen.

\begin{figure*}
\begin{picture}(400,350)
\put(-50,0){\includegraphics[height=350pt]{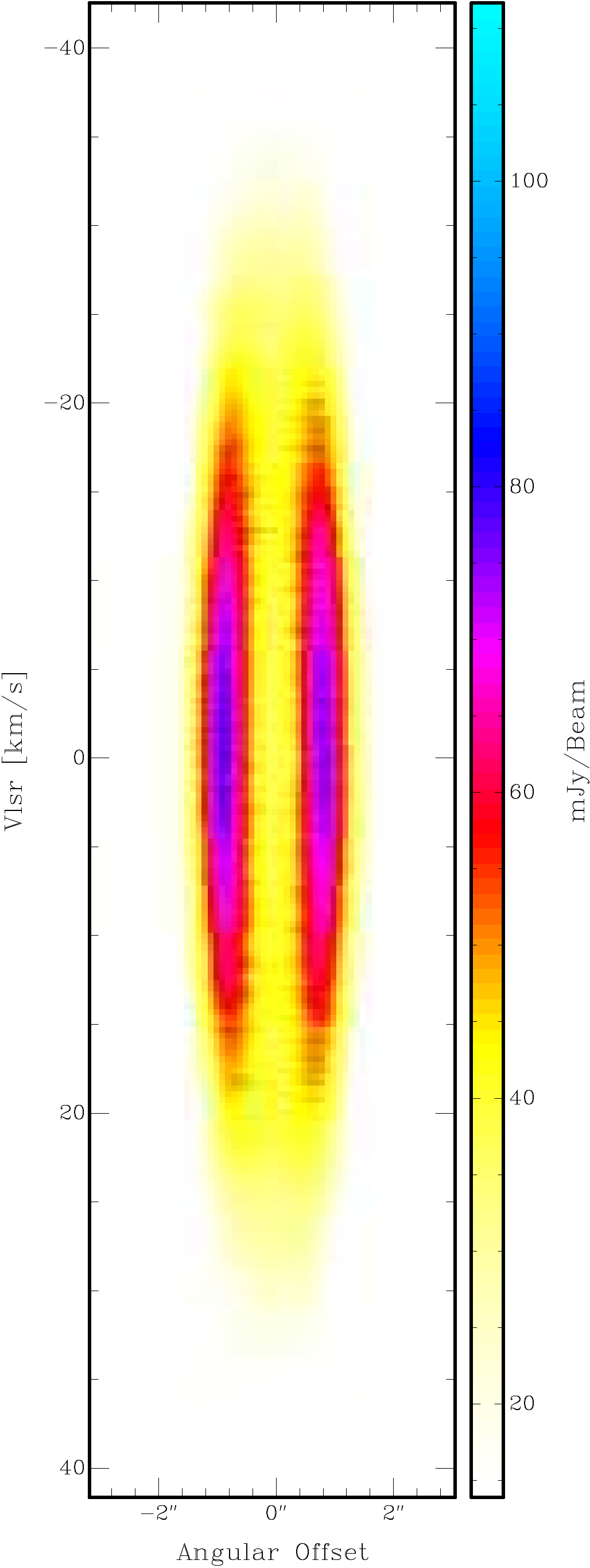}}
\put(110,250){\includegraphics[height=350pt,angle=-90]{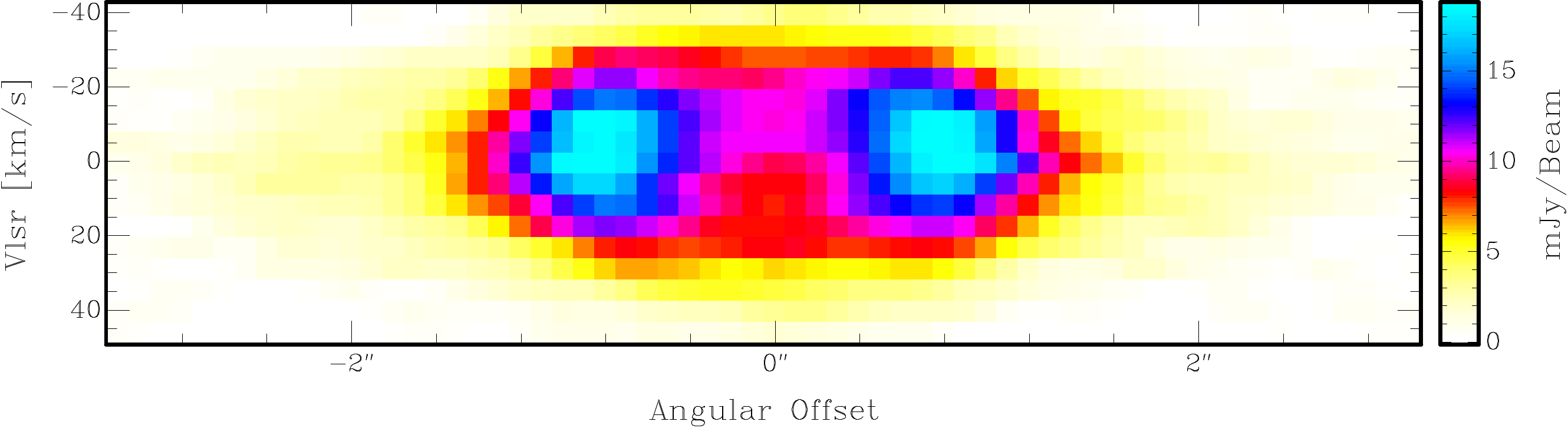}}
\end{picture}
\caption{Position-velocity (PV) diagrams for the H26$\alpha$
(left) and H68$\alpha$ (right) cubes along the cut shown in Figure~\ref{fig:h68a}
for the 0.3$\arcsec$ resolution simulated observations.}
\label{fig:pv}
\end{figure*}

The same information can be seen in the position-velocity (PV)
diagrams in Figure~\ref{fig:pv}. The H26$\alpha$ PV diagram
is nearly symmetric about the systemic
velocity (0\,km/s). However, the H68$\alpha$ is clearly asymmetric and
blue-shifted with respect to the systemic velocity.

%------------------------------
%------------------------------

%------------------------------------------------------------
\section{Requirements to observe asymmetric RRL profiles in \hii\ regions}
\label{sec:preditc_conditions}

RRL line profile asymmetries are a potentially powerful probe of
systematic motions (infall and outflow) in ionised gas. The results
from the simulated observations above show that being able to resolve
the optically thick region is crucial for any observations hoping to
detect these asymmetries. Using a basic \hii\ region model and making
some simple assumptions we now try to predict the conditions necessary
to observe asymmetric RRL profiles for a range of \hii\ region
properties.

We start with a basic model of an \hii\ region as a sphere with
uniform electron number density $\nel$, temperature $T$ and
radius $R$ which is at a distance $D$. While such a model is undoubtedly
oversimplified, these physical properties should be thought of as
similar to the beam-averaged values reported by observers for
\hii\ regions which probably have either multiple density and
temperature components or temperature and density gradients.

The critical physical scale for observing the RRL asymmetries is the
radius at which the emission becomes optically thick. We denote this
radius $R_\mathrm{thick}$. With gas at a uniform electron density and
temperature, it is trivial to solve the one-dimensional radiative transfer equation
and calculate the path length along the line of sight for which the
emission optical depth equals unity for a given RRL frequency. We
define $2\times R_\mathrm{thick}$ to be the path length for which a photon of a
given frequency, emitted from the back of the \hii\ region would be
reabsorbed before reaching the front edge of the \hii\ region---i.e. the path
length for which $\tau = 1$.

\begin{figure}
\centerline{\includegraphics[height=170pt]{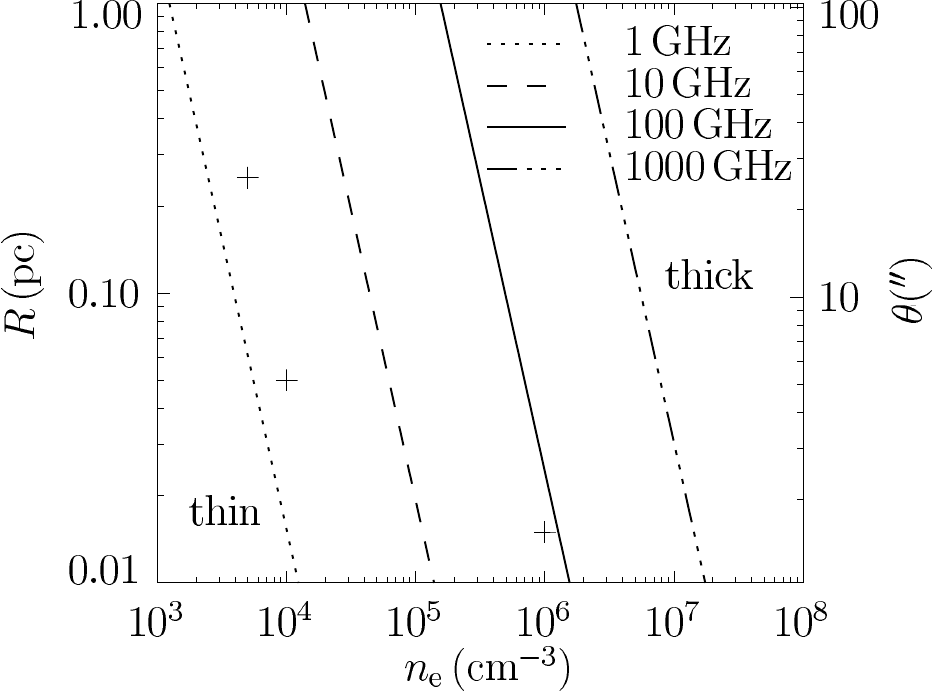}}
\caption{Threshold radius $R_\mathrm{thick}$ (and corresponding angular scale $\theta$ at a
distance of $2$\,kpc) at which the emission becomes optically thick
as function of electron number denity $\nel$ and for frequencies of $1$, $10$, $100$ and $1000$\,GHz.
The crosses mark the regimes (from left to right) of compact, ultracompact and hypercompact \hii\ regions, respectively.}
\label{fig:rnelplot}
\end{figure}

Figure~\ref{fig:rnelplot} shows the dependence of $R_\mathrm{thick}$ on observing
frequency as a function of $\nel$ for \hii\ regions with
$T = 10^4$\,K. For reference, the angular size of a source with radius
$R_\mathrm{thick}$ at a distance of 2\,kpc is given on the right-hand
vertical axis. Overlayed on this plot are the typical size and density
of compact, ultracompact and hypercompact \hii\ regions (following \citealt{kurtz05}), showing the
regimes at which the emission is optically thick or thin. For the
densest hypercompact \hii\ regions, frequencies larger than 100\,GHz are
required to probe the optically thin emission.

\begin{figure}
\centerline{\includegraphics[height=170pt]{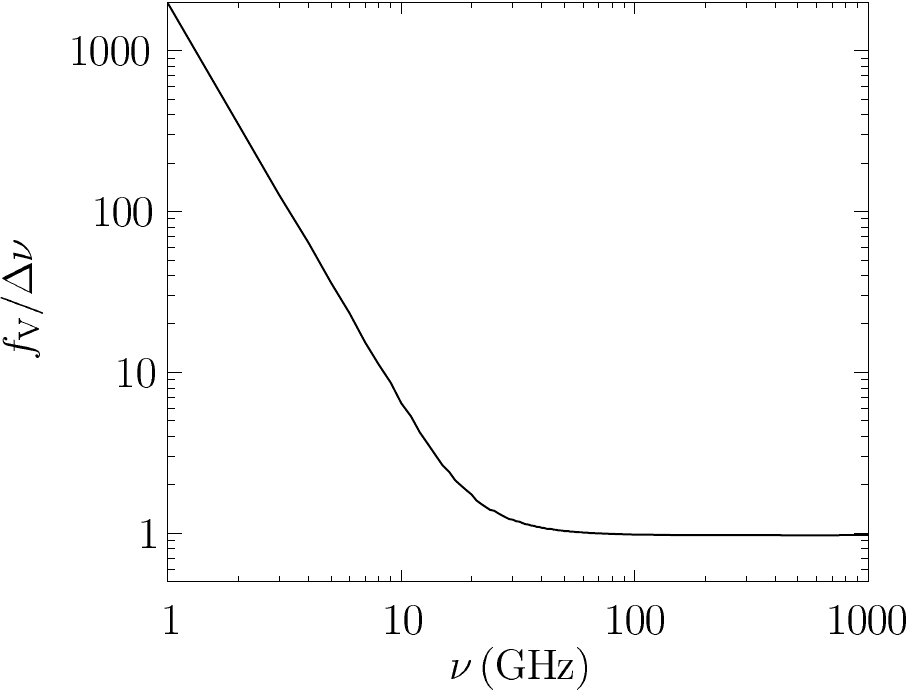}}
\caption{Ratio of the Voigt profile FWHM $f_\mathrm{V}$ and the Doppler shift $\Delta \nu$
as function of line frequency $\nu$ for $\nel = 10^6\,$cm$^{-3}$ and $\Delta v = 11\,$km/s.
Preasure broadening dominates at low frequencies but becomes negligible above $\sim 100\,$GHz.}
\label{fig:broadening}
\end{figure}

\begin{figure}
\centerline{\includegraphics[height=170pt]{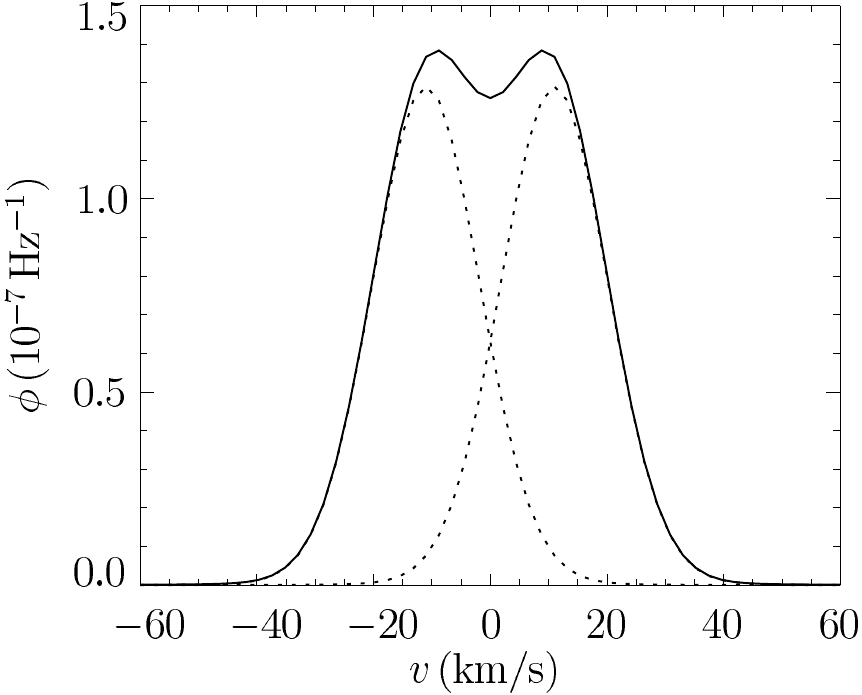}}
\caption{The red- and blue-shifted line profiles and their sum for the same conditions
as in Figure~\ref{fig:broadening} at $\nu = 100\,$GHz. The individual components can
be separated.}
\label{fig:profiles}
\end{figure}

To distinguish the regime at which the intrinsic line width is too
broad to spectrally resolve the line asymmetries we need to assume
some typical velocity structure in the gas. The sound speed in ionized
gas, $\cs$, seems a sensible choice here. For example, a region
undergoing pressure-driven expansion will initially expand at the sound
speed. Therefore, the most red-shifted and blue-shifted emission would
be separated in velocity by $2 \times \cs$. Since the temperature inside an \hii\ region
is approximately $T = 10^4\,$K everywhere, the corresponding sound speed is
about $\cs = 9$\,km/s. As outlined in the Section~\ref{sec:physrecomb}, the intrinsic RRL line
profile is a Voigt profile resulting from the contributions of a
Gaussian thermally-broadened component and a Lorentzian
pressure-broadened component. The width of the Voigt profile can be
characterized with its full width at half maximum (FWHM) $f_\mathrm{V}$.
To see under which conditions the red- and blue-shifted components
can be observed individually, $f_\mathrm{V}$ must be compared to the
Doppler shift due to the systematic velocity difference $\Delta v$,
\begin{equation}
\label{eq:deltanu}
\Delta \nu = 2 \frac{\Delta v}{c} \nu \text{,}
\end{equation}
where $c$ is the speed of light and the factor $2$ comes from the fact
that both the red- and blue-shifted components have a velocity $\Delta v$, but
with opposite signs.
This leads to the requirement that $\Delta \nu \geq f_\mathrm{V}$.
To be definite, we take a typical number
density of a hypercompact \hii\ region of $\nel = 10^6\,$cm$^{-3}$
and a velocity slightly greater than the sound speed, $\Delta v = 11\,$km/s.
Figure~\ref{fig:broadening} shows the ratio $f_\mathrm{V} / \Delta \nu$
as function of the line frequency $\nu$.
The profile width $f_\mathrm{V}$ increases dramatically at low
frequencies due to the pressure broadening term but approaches the
thermal linewidth at high frequencies. The red- and blue-shifted
components as well as their sum is shown in Figure~\ref{fig:profiles}
at a frequency of $100\,$GHz, where the two peaks can be individually
resolved.

The figures demonstrate that pressure broadening will not
be problematic for ALMA observations even of hypercompact \hii\ regions.
On the other hand, the systematic gas motion must be at least as large
as the sound speed to be detectable due to the thermal broadening.
Since both the thermal line width~\eqref{eq:sigma}
and the spectral separation~\eqref{eq:deltanu} scale linearly with
frequency $\nu$, sub-thermal motion within the \hii\ region cannot
be resolved independent of frequency. The threshold frequency where
the two peaks in the line profile can be separated is larger the smaller
the systematic velocity difference and the denser the \hii\ region is.
Our example demonstrates that even slightly
super-sonic velocities can be separated in most ALMA bands for typical \hii\ region densities provided that
the spectral resolution is significantly smaller than $\Delta \nu$. 
Of course, the detailed line profile shape and the ability for ALMA to
resolve multiple velocity components will depend on many additional factors
(e.g. signal to noise ratio of each component, relative strength of each component etc.).

\begin{figure}
\centerline{\includegraphics[height=170pt]{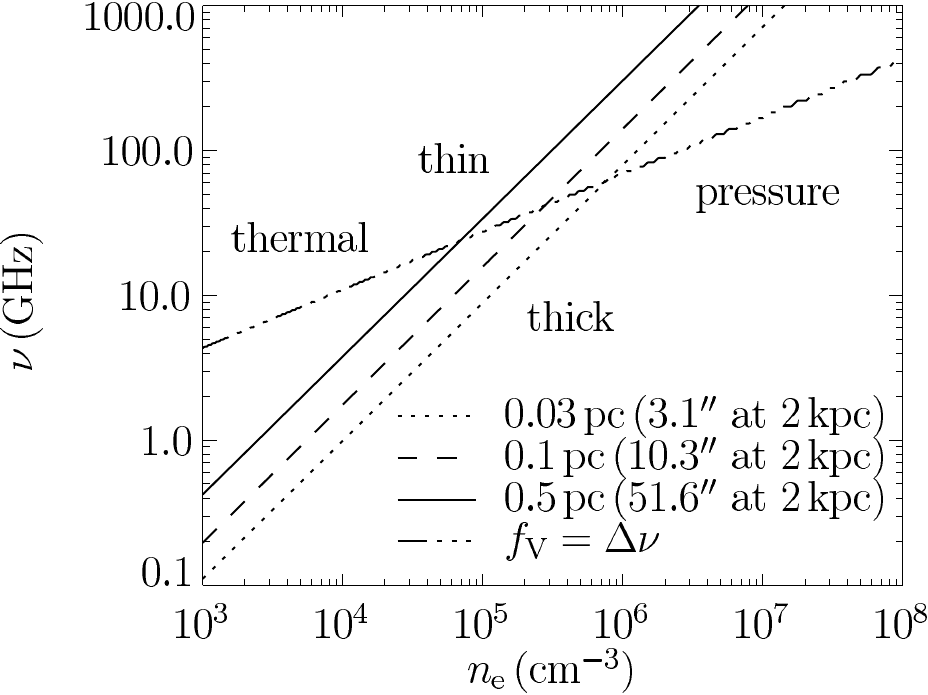}}
\caption{Threshold frequency for the optically thick-thin transition for \hii\ regions
with diameters 0.03, 0.1 and 0.5\,pc (dotted, dashed and solid, respectively) and threshold frequency
where $f_\mathrm{V} = \Delta \nu$ (dot-dash) as function of electron number density $\nel$ inside
the \hii\ region. Line asymmetries can be expected in the upper right corner of the diagram.}
\label{fig:nenu}
\end{figure}

Figure~\ref{fig:nenu} summarises these results and those in Figure~\ref{fig:rnelplot}. It
provides the predicted observational conditions required in order to
detect RRL profile asymmetries from an \hii\ region with
$T = 10^4$\,K. The dotted, dashed and solid lines show
the threshold frequency for which \hii\ regions with diameters of
0.03, 0.1 and 0.5\,pc (matching the typical sizes of hypercompact, ultracompact and
compact \hii\ regions, respectively \citep{kurtz05}) at the given
density become optically thick for free-free radiation. The angular
scales these diameters correspond to for a source at 2\,kpc are shown in
parentheses. For an \hii\ region of a given density and diameter,
emission at frequencies below the line will be optically thick, while
emission above the line will be optically thin.

The dot-dash line in Figure~\ref{fig:nenu} shows the frequency at which
$f_\mathrm{V} = \Delta \nu$ for $\Delta v = 11\,$km/s and
an \hii\ region of a given density. As
described above, we define this as a conservative lower limit to the
intrinsic linewidth at which it is possible to resolve line profile
asymmetries. It is only at frequencies above this line, that the
intrinsic linewidth would be small enough to detect the asymmetries.

This important point from Figure~\ref{fig:nenu} is that only the upper-right hand
corner of the plot (above the dot-dash line and below the solid,
dashed and dotted line) satisfies the conditions for the line profile
asymmetries to be detected. It them becomes clear why strong RRL line
profile asymmetries have not been reported much before. The
asymmetries have been difficult to detect due to a combination of the
observational limits imposed by previous telescopes, in particular the
brightness-temperature sensitivity, angular resolution and spectral
resolution.

It is clear that single-dish telescopes do not have sufficient
resolution to resolve the optically thick component. Only
interferometers pass the angular resolution criteria. At cm
wavelengths the VLA has been the premier facility in terms of angular
resolution and surface brightness sensitivity. Indeed, much of the
pioneering recombination line work was done with the VLA several
decades ago at frequencies smaller than 20\,GHz \citep[e.g.][]{garayetal86,depreeetal95b}. However, as
shown in Figure~\ref{fig:broadening} the pressure broadening at these low frequencies is
so large that it may have dominated the line profile and masked any
asymmetries. Also, the limitations imposed by the previous correlator made it difficult
to find spectral setups with simultaneously large enough bandwidth to fit the
very broad lines (larger than 100\,km/s) of the high-$n$ transitions, while at
the same time providing the spectral resolution to resolve line asymmetries.
Only having a few resolution elements across the line profile would make it
difficult to identify asymmetries. These correlator restrictions have
been removed with the upgrade of the VLA to its reincarnation as the
EVLA. EVLA observations of RRLs towards dense \hii\ regions at high
angular/velocity resolution and sensitivity may be able to detect the
profile asymmetries.

Higher frequency interferometers (e.g. SMA, CARMA, PdBI) do not suffer
this bandwidth/velocity resolution problem. Indeed, in going to lower
$n$, the pressure broadening drops off rapidly (Equation~\eqref{eq:delta}, Figure~\ref{fig:broadening}).
However, for a number of reasons the surface-brightness
sensitivities of these interferometers is lower than that of the
VLA. The simulated observations above show a lot of integration time
would be required to achieve high enough resolution and, at the same
time, sensitivity to detect asymmetric profiles.

In summary, it is only with the order of magnitude increase in
surface-brightness sensitivity, higher angular resolution and higher
frequencies accessible with ALMA and the improved correlator with the
EVLA that these may become routinely observable.

\section{Summary}
\label{sec:summary}

We have presented a discussion of the capabilities for ALMA to observe
hydrogen recombination lines and of their physical properties. An important
difference between ALMA lines and lines at cm-wavelengths is that the level
populations of the former lines are farther away from LTE conditions, which
gives rise to the formation of asymmetric line profiles. We have
investigated necessary conditions to observe such profiles with ALMA and
presented synthetic ALMA and EVLA observations of simple model \hii\ regions.
We find that the asymmetric line profiles can provide useful information
about systematic gas motion, such as infall or outflow, within the \hii\ region
that was previously unaccessible. Radiative transfer modeling, e.g.
with the RADMC-3D implementation of recombination lines presented here,
will be necessary to distinguish non-LTE effects from geometric effects
due to asymmetries in the \hii\ region and to extract the desired information
from the observations. Interested readers can obtain the code
upon request from the first author.

\section*{Acknowledgements}

We thank Roberto Galv{\'a}n-Madrid, Chris De Pree and Stan Kurtz for 
helpful comments and stimulating discussions. We also thank the anonymous referee
for very useful comments that helped to improve the paper.
T.P. acknowledges financial support as a Fellow of the
Baden-W\"{u}rttemberg Stiftung funded by their program International
Collaboration II (grant P-LS-SPII/18) and through SNF grant
200020\textunderscore 137896. S.N.L. acknowledges the research leading to
these results has received funding from the European Community's
Seventh Framework Programme (/FP7/2007-2013/) under grant agreement No.
229517.

%------------------------------------------------------------
\appendix

\section{Symmetry of Line Profiles}
\label{sec:symm}

In this Section, we investigate under which conditions observed line profiles from
symmetric \hii\ regions are symmetric. The result will be that line profiles from isothermal, symmetric \hii\ regions under
LTE conditions will always be symmetric, independent of optical depth effects, whereas
asymmetric line profiles are typical for \hii\ regions in non-LTE states. This was previously noticed
by \citet{rodr82}, but we want to make the original argument clearer.
To simplify the calculation we will neglect the continuum opacity in this problem. The continuum opacity varies
only mildly across the line profile compared to the line opacity, and we are interested here in true line profile
asymmetries, not just a profile on top of an asymmetric continuum background. 

\begin{figure}
\centerline{\includegraphics{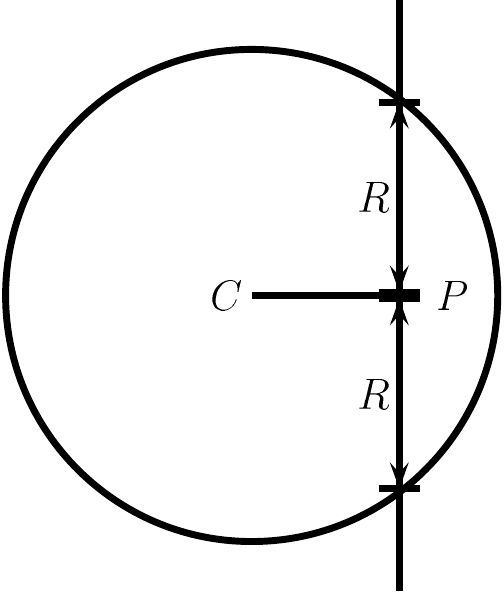}}
\caption{A ray traversing a spherically symmetric \hii\ region. A radial line from the center $C$
cuts the ray at right angle at a point $P$. The distance along the ray from $P$ to the edge of the \hii\ region
is $R$. Because of the spherical symmetry, all physical quantities along the ray are mirror-symmetric with
respect to the point $P$.}
\label{fig:sketch}
\end{figure}

We consider a ray passing through an isothermal, spherically symmetric \hii\ region (see Figure~\ref{fig:sketch}).
Because of the symmetry, all physical quantities along the ray will be mirror-symmetric with respect to the point where
a radial line from the center of the \hii\ region $C$ cuts the ray at right angle. Let this point $P$ be the origin of our
coordinate system, and let $R$ be the distance from $P$ to the edge of the \hii\ region along the ray, then we can
write the total optical depth as
\begin{equation}
\label{eq:taup}
\tau_{\nu_0 + \nu'} = \int_{-R}^R \alpha_{\nu_0 + \nu',\mathrm{L}}(r')\,\mathrm{d}r' .
\end{equation}
Because of the mirror symmetry we have $v(-r') = -v(r')$, and
with the symmetry of the profile function~\eqref{eq:voigtprof} we find
\begin{equation}
\label{eq:arel}
\alpha_{\nu_0 + \nu',\mathrm{L}}(-r')
= \alpha_{\nu_0 - \nu',\mathrm{L}}(r') .
\end{equation}
By decomposing the integral~\eqref{eq:taup} and using relation~\eqref{eq:arel}, we find
\begin{equation}
\label{eq:taueq}
\tau_{\nu_0 + \nu'} = \int_0^R (\alpha_{\nu_0 + \nu',\mathrm{L}}(r') + \alpha_{\nu_0 - \nu',\mathrm{L}}(r'))\,\mathrm{d}r' .
\end{equation}
This expression is evidently symmetric in $\nu'$, and hence we have $\tau_{\nu_0 + \nu'} = \tau_{\nu_0 - \nu'}$, the
total optical depth of the ray is symmetric. This result is independent of any LTE assumptions.

The above argument demonstrates that the optical depth is only symmetric after integration
over the whole region. For positions inside the \hii\ region, say between $P$ and the observer, the two contributions in Equation~\eqref{eq:taueq}
do not cancel because the term from the rear side is already included while the corresponding contribution
from the front side is not, and a frequency dependence remains.

In LTE, the source function of an isothermal \hii\ region is simply $S_\nu = B_\nu(T)$ everywhere,
and thus the integral~\eqref{eq:rtint} reduces to
\begin{equation}
I_\nu(\tau_\nu) = B_\nu(T) (1 - \mathrm{e}^{-\tau_\nu}) .
\end{equation}
Since the total optical depth $\tau_\nu$ is symmetric, so is the intensity $I_\nu$ at the observer,
neglecting the small variation of $B_\nu(T)$ over the line profile.

In non-LTE, however, the source function~\eqref{eq:sourfun} is
\begin{equation}
S_\nu = B_\nu(T) \frac{b_m}{b_n \beta_{n,m}} ,
\end{equation}
where we have again neglected the continuum opacity. Thus, the intensity is now given by the integral
\begin{equation}
I_\nu(\tau_\nu) = \mathrm{e}^{-\tau_\nu} B_\nu(T) \int_0^{\tau_\nu} \mathrm{e}^{\tau_\nu'} \frac{b_m(\tau_\nu')}
{b_n(\tau_\nu') \beta_{n,m}(\tau_\nu')}\,\mathrm{d}\tau_\nu' .
\end{equation}
There is no reason why this expression should still be symmetric. In fact, as noted above,
$\tau_\nu$ is only symmetric after integrating over the whole ray. The expression $\mathrm{e}^{\tau_\nu'}$ in the
integral will not necessarily be symmetric, leading to a non-symmetric value of the integral,
and hence the non-LTE departure coefficients break the symmetry
of the line profile\footnote{Although we have neglected the continuum opacity to simplify the mathematics,
it is clear that the effect will grow with $\tau_\nu'$ and thus be largest in optically thick regions.}.
Our implementation tests (see Section~\ref{sec:linshap}) demonstrate that this effect can
be surprisingly big, proving the necessity of using non-LTE conditions in simulating recombination line observations.

%------------------------------------------------------------
\section{Tests}
\label{sec:tests}

We verify our implementation by a series of tests conducted by \citet{vinvalhug79}.
All radiative transfer tests are run with free-free radiation and recombination lines combined.
After the radiative transfer, we use a polynomial first-order fit to subtract the continuum contribution
from the images.

The tests are carried out for three different density profiles, which we denote as model 1 to 3.
For all three models we assume a constant gas temperature of $T = 10^4$\,K and a microturbulent velocity
of $\xi = 15$\,km\,s$^{-1}$. The models differ in the radial density profiles, which are defined
as follows:
\begin{itemize}
\item model 1:
\begin{equation}
\label{eq:model1}
\nel =
\begin{cases}
28 \times \left(\frac{\displaystyle r}{\displaystyle \mathrm{pc}}\right)^{-2}\,\mathrm{cm}^{-3} & 0.02\,\mathrm{pc} \leq r \leq 0.1\,\mathrm{pc}\\
0 & \mathrm{otherwise} ,
\end{cases}
\end{equation}
\item model 2:
\begin{equation}
\nel =
\begin{cases}
10^5 \times \exp(-700\, (r / \mathrm{pc})^{2})\,\mathrm{cm}^{-3}  & r \leq 0.1\,\mathrm{pc}\\
0 & \mathrm{otherwise} ,
\end{cases}
\end{equation}
\item model 3:
\begin{equation}
\nel =
\begin{cases}
5 \times 10^4\,\mathrm{cm}^{-3} & r \leq 0.24\,\mathrm{pc}\\
0 & \mathrm{otherwise} .
\end{cases}
\end{equation}
\end{itemize}
We also consider models with expanding and contracting gas velocities. These models are identical
to model~1, but additionally include a radial gas velocity $v_r$ according to the following prescription:
\begin{itemize}
\item model 4:
\begin{equation}
v_r = 80\,\mathrm{km\,s}^{-1} ,
\end{equation}
\item model 5:
\begin{equation}
v_r = -80\,\mathrm{km\,s}^{-1} ,
\end{equation}
\item model 6:
\begin{equation}
\label{eq:model6}
v_r = 25 \times \left(1 - \frac{r}{0.1\,\mathrm{pc}}\right)\,\mathrm{km\,s}^{-1} ,
\end{equation}
\item model 7:
\begin{equation}
v_r = -25 \times \left(1 - \frac{r}{0.1\,\mathrm{pc}}\right)\,\mathrm{km\,s}^{-1} .
\end{equation}
\end{itemize}
There are three major differences between our implementation in RADMC-3D and the calculations by \citet{vinvalhug79}.
First, \citet{vinvalhug79} used an adaptive step size to integrate the radiative transfer problem~\eqref{eq:rt}.
In our calculations, unless otherwise noted, we always map the model setups to a homogeneous, Cartesian grid with
$100^3$ grid points and linear dimension $0.5\,$pc. Second, \citet{vinvalhug79} determined the line intensity
by comparing radiative transfer calculations of the full emission and absorption coefficients,
$j_{\nu,\mathrm{C}} + j_{\nu,\mathrm{L}}$ and $\alpha_{\nu,\mathrm{C}} + \alpha_{\nu,\mathrm{L}}$, respectively,
with reduced calculations taking only continuum contributions by $j_{\nu,\mathrm{C}}$ and $\alpha_{\nu,\mathrm{C}}$
into account. Since this procedure is unavailable to observers, we prefer to work with the full radiative
transfer problem exclusively and determine the line intensities by continuum subtraction. Third, \citet{vinvalhug79}
include not only hydrogen but also helium in their models. This leads to a marginally larger electron number density
compared to a the pure hydrogen case as in our models, and hence can produce slightly enhanced line intensities.

\subsection{Line Strength as Function of Frequency}

\begin{figure}
\centerline{\includegraphics[height=170pt]{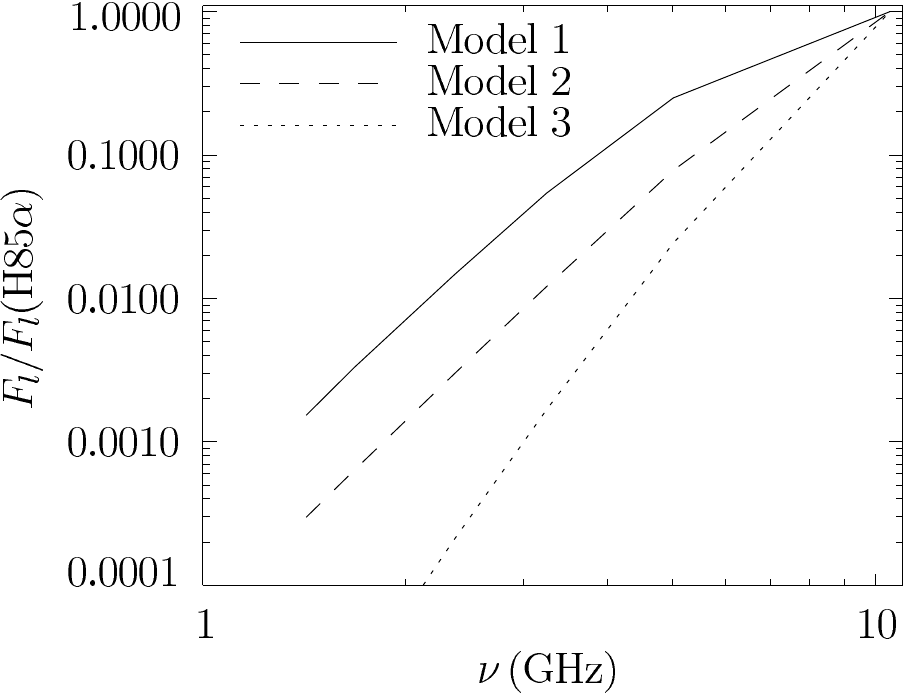}}
\caption{The peak line flux relative to H85$\alpha$ for the recombination lines H109$\alpha$, H126$\alpha$,
H140$\alpha$, H157$\alpha$ and H166$\alpha$ for models~1, 2 and 3. Compare Figure 3 in \citet{vinvalhug79}.}
\label{fig:fig3}
\end{figure}

To study the variation of the line strength as function of the transition frequency, we show in
Figure~\ref{fig:fig3} the peak line flux relative to the H85$\alpha$ line for the
recombination lines H109$\alpha$, H126$\alpha$, H140$\alpha$, H157$\alpha$ and H166$\alpha$
for the models~1, 2 and 3. These transitions span almost one order of magnitude in frequency, and
the resulting peak line flux varies by more than four orders of magnitude. The line strength
increases monotonically with decreasing principal quantum number number and is always the largest for
model~1 and the smallest for model~3 for a certain transition, with model~2 lying in between. The small differences between
Figure~\ref{fig:fig3} and Figure~3 in \citet{vinvalhug79} are likely the result of different methods
for subtracting the continuum since the peaks of the higher transition lines are very weak.
 
\subsection{Line Strength and Width as Function of Order}
\label{sec:linstrwid}

\begin{figure}
\centerline{\includegraphics[height=170pt]{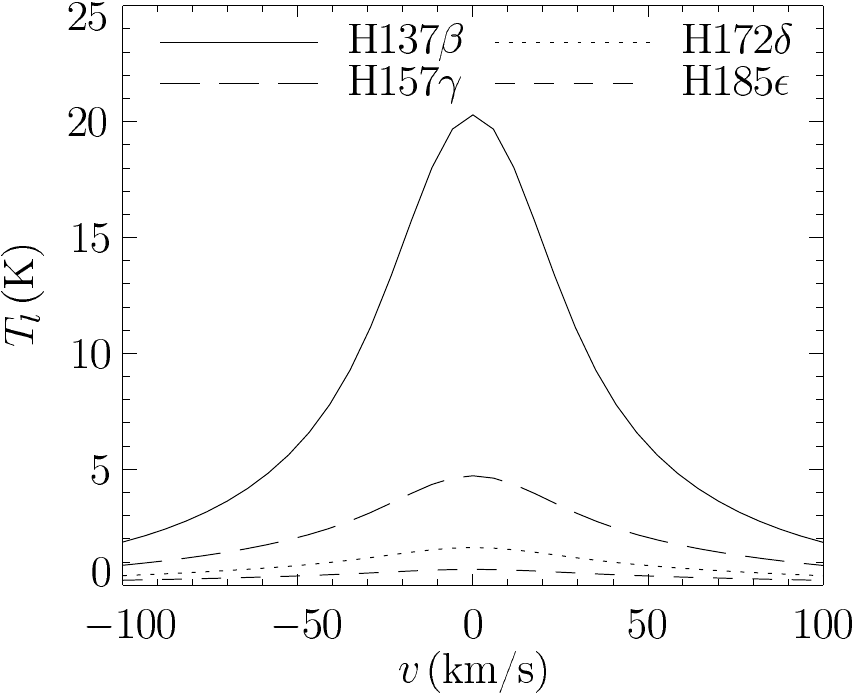}}
\caption{Line profiles for rays through the center of the sphere in model~1. The line temperatures are
calculated for H137$\beta$, H157$\gamma$, H172$\delta$ and H185$\epsilon$ lines.
Compare Figure 2c in \citet{vinvalhug79}.}
\label{fig:fig2c}
\end{figure}

Figure~\ref{fig:fig2c} shows the line profiles along a ray through the center of the sphere of model~1
for four different recombination lines near 5~GHz with increasing transition order $\Delta n$
(H137$\beta$, H157$\gamma$, H172$\delta$ and H185$\epsilon$). The peak line temperatures decrease
and the line width increases with increasing order of the transition. There are small deviations $\lesssim 15\%$
from Figure~2c in \citet{vinvalhug79} in the height of the line peaks. This could in principle be a result
of insufficient spatial resolution near the density peak in model~1 on our homogeneous grid. To test this,
we have run a radiative transfer calculation with adaptive mesh refinement for these lines
(see Section~\ref{sec:amrtest} for further discussion). To resolve
the density peak in the $r^{-2}$-profile of model~1, we introduce a Jeans-like refinement criterion.
We require that the cell size $\mathrm{d}x$ and the electron number density $\nel$ always satisfy
the relation
\begin{equation}
\label{eq:jeansref}
6 \times \frac{\mathrm{d}x}{\mathrm{pc}} < \left(5 \times \frac{\nel}{\mathrm{cm}^{-3}}\right)^{-1/2} .
\end{equation}
If this condition is not met, we split the cell under consideration into eight smaller cells with
half the linear dimension. This procedure is iterated until equation~\eqref{eq:jeansref} is satisfied
everywhere. This leads to the creation of five refinement levels on top of the $100^3$ base grid.
The total adaptive mesh consists of 27,282,088 cells, which is more than 27 times the number of cells
in the homogeneous case. However, the line peaks are amplified only very mildly, so that the discrepancy
to the results of \citet{vinvalhug79} remains. The same holds for a spherically symmetric mesh with
a logarithmic radial coordinate axis (see Section~\ref{sec:amrtest}), so that the integration method is
unlikely to be the reason.

\begin{figure}
\centerline{\includegraphics[height=170pt]{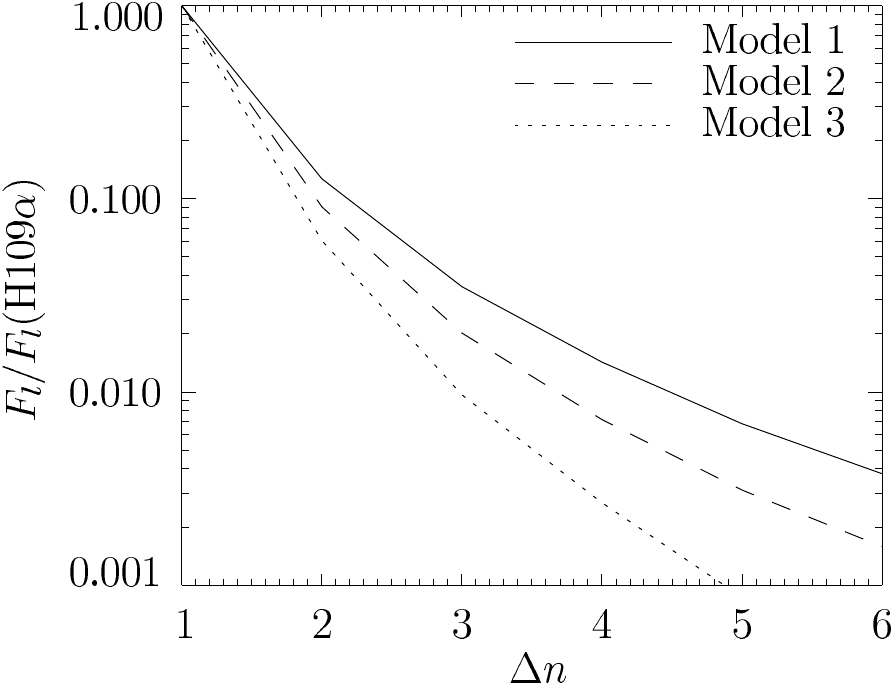}}
\caption{Source-integrated peak line fluxes for H137$\beta$, H157$\gamma$, H172$\delta$, H185$\epsilon$ and H196$\zeta$
relative to H109$\alpha$ for models~1, 2 and 3. Compare Figure~4a in \citet{vinvalhug79}.}
\label{fig:fig4a}
\end{figure}

Figure~\ref{fig:fig4a} displays source-integrated peak line fluxes relative to H109$\alpha$ for some lines
near 6\,cm (H137$\beta$, H157$\gamma$, H172$\delta$, H185$\epsilon$ and H196$\zeta$) as function
of the transition order $\Delta n$ for models~1, 2 and 3. The peak line flux monotonically decreases
with increasing $\Delta n$. The peak line flux is the largest for model~1, the second-largest for model~2
and the smallest for model~3 for a fixed $\Delta n$. The small deviations from Figure~4a in \citet{vinvalhug79}
are most probably introduced by the continuum subtraction since again the higher-order lines are
very weak compared to the continuum.

\begin{figure}
\centerline{\includegraphics[height=170pt]{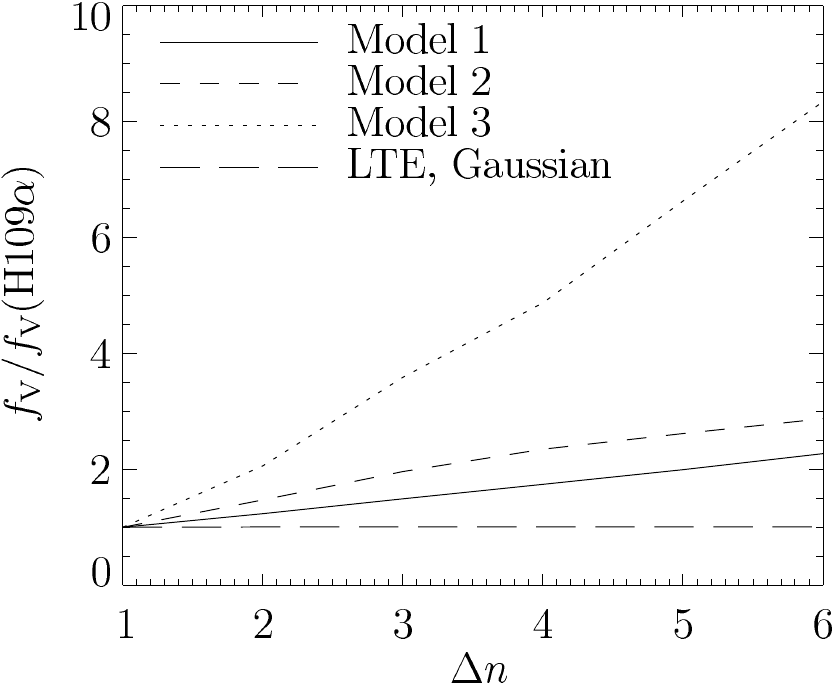}}
\caption{Linewidth for the same lines as in Figure~\ref{fig:fig4a} and a control calculation of model~1
with LTE occupation numbers and Gaussian line profiles. Compare Figure~4b in \citet{vinvalhug79}.}
\label{fig:fig4b}
\end{figure}

Figure~\ref{fig:fig4b} shows the line width for the same source-integrated lines as in Figure~\ref{fig:fig4a}.
We use the full width at half maximum (FWHM) as a measure of the line width. To determine the FWHM of the lines, we
fit a Voigt profile to the observed line profiles and extract the Doppler width $\sigma$ and the Lorentz parameter $\delta$.
The FWHM of the Voigt profile can be computed via the FWHM of the Gaussian profile $f_\mathrm{G} = 2 \sqrt{2 \ln 2} \sigma$
and the Lorentzian profile $f_\mathrm{L} = 2 \delta$ following \citet{olivlong77} as
\begin{equation}
f_\mathrm{V} = \frac{1}{2} \left(C_1 f_\mathrm{L} + \sqrt{C_2 f_\mathrm{L}^2 + 4 C_3 f_\mathrm{G}^2}\right)
\end{equation}
with the parameters $C_1 = 1.0692$, $C_2 = 0.86639$ and $C_3 = 1.0$.

The line width in Figure~\ref{fig:fig4b} increases monotonically with the order of the transition. This is because
the principal quantum number of the transition also increases, and following Equation~\eqref{eq:delta} pressure
broadening becomes more important for higher quantum numbers. We also show a calculation with LTE occupation
numbers and a pure Gaussian line profile for model~1 for comparison. While the linewidth for models~1 and 2
agree relatively well with Figure~4b in \citet{vinvalhug79}, the linewidth for model~3 is larger by a factor
of about two in our computation for high-order transitions. This is probably because \citet{vinvalhug79} use a different
fitting formula for the $\delta$ parameter of the Lorentzian profile.

\begin{figure}
\centerline{\includegraphics[height=170pt]{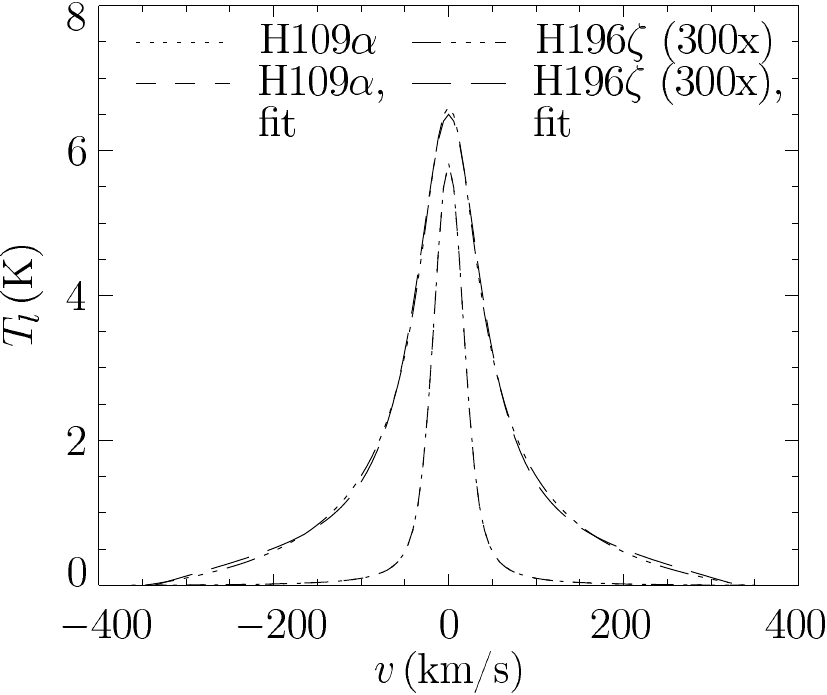}}
\caption{Source-integrated line temperatures for the H109$\alpha$ line and H196$\zeta$ line (multiplied by 300)
for model~1. The plot shows both the measured profiles and the Voigt fits to the profiles. The fits represent
the data very well.}
\label{fig:samplprof}
\end{figure}

To demonstrate the goodness of the Voigt profile fit to the measured line profiles as well as the
quality of the derived linewidth parameters we show the measured line temperatures for model~1 for
the lines H109$\alpha$ ($\Delta n = 1$) and H196$\zeta$ ($\Delta n = 6$) in Figure~\ref{fig:samplprof}.
The line temperature for the H196$\zeta$ is multiplied by a factor of 300 to allow a direct comparison
to the H109$\alpha$ line. The Voigt fits are very good, which verifies that the derived
Voigt profile parameters $\sigma$ and $\delta$ have converged.

\subsection{Line Shape as Indicator of Gas Motion}
\label{sec:linshap}

\begin{figure}
\centerline{\includegraphics[height=170pt]{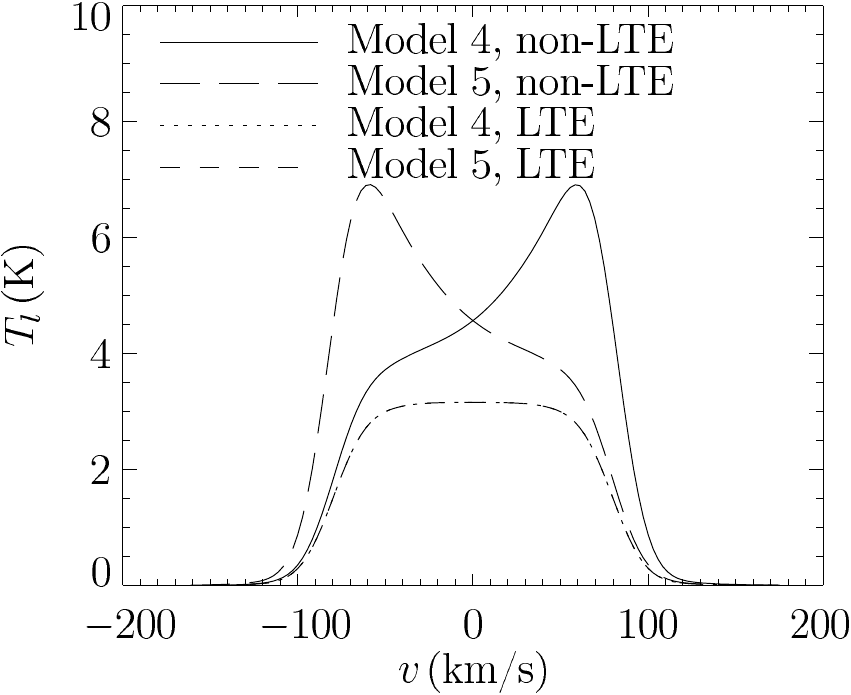}}
\caption{Source-integrated line temperature of H85$\alpha$ for models~4 and 5 in LTE and non-LTE. The LTE
line profiles for expanding (model~4) and contracting (model~5) gas motion are identical and symmetric, whereas the
non-LTE line profiles are clearly distinguishable and asymmetric. Compare Figure~5 in \citet{vinvalhug79}.}
\label{fig:fig5}
\end{figure}

To investigate the effect of radial gas motion on the line shape, we consider H85$\alpha$ observations
of models~4 to 7. As explained in Appendix~\ref{sec:symm}, such line profiles must always be symmetric with
respect to the rest frequency in LTE calculations. Figure~\ref{fig:fig5} shows the source-integrated
line temperature for the case of constant radial expansion (model~4) and contraction (model~5). Indeed,
the LTE profiles are symmetric and cannot distinguish between expanding and contracting motion.
The non-LTE profiles, however, are highly asymmetric.
\citet{vinvalhug79} explained the asymmetric line profile as a result of optical depth effects
(see also \citet{escalanteetal89}).
As explicated in Appendix~\ref{sec:symm}, it is crucial to carry out the calculation under non-LTE
conditions to see this effect.

\begin{figure}
\centerline{\includegraphics[height=170pt]{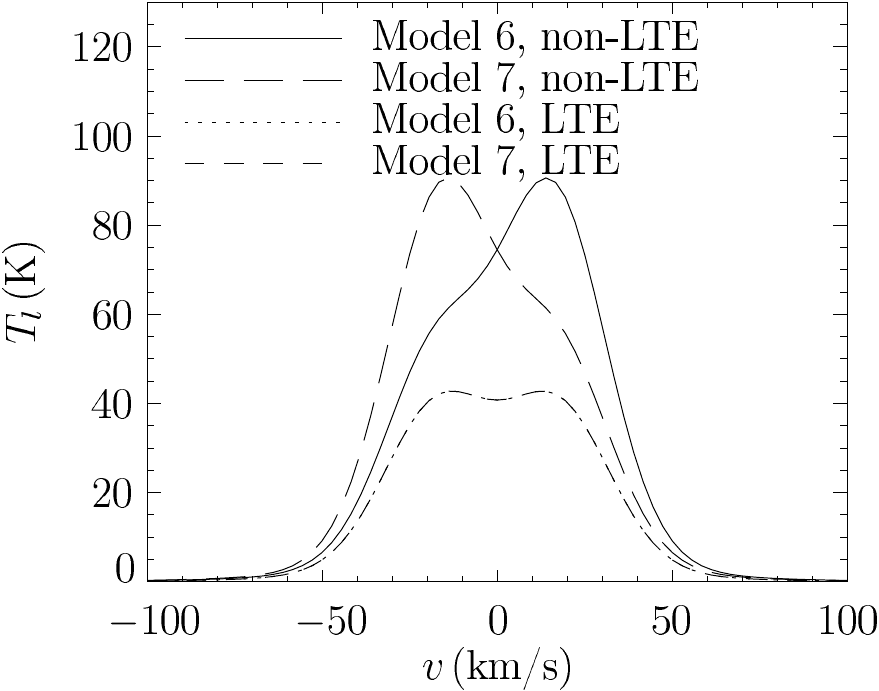}}
\caption{Line temperature of H85$\alpha$ for a line of sight through the center of the sphere for
models~6 and 7 in LTE and non-LTE. The LTE line profiles are symmetric and identical for model~6 and model~7,
while the non-LTE profiles differ because of their asymmetry. Compare Figure 6 in \citet{vinvalhug79}.}
\label{fig:fig6}
\end{figure}

Figure~\ref{fig:fig6} demonstrates the same mechanism for models~6 and 7, this time for a
line of sight through the center of the sphere. The strength of the effect varies with
spatial position and is strongest for the line of sight shown. It is interesting that such
highly asymmetric line profiles can be produced simply through non-LTE conditions, without
any asymmetries in the temperature, density or velocity distribution.

\subsection{Maser Amplification}

In principle, the non-LTE conditions within \hii\ regions can give rise to maser amplification of the
recombination line emission \citep{gorsor02}.
To test the ability of our code to model maser amplification, we conduct the maser test problem of
\citet{vinvalhug79}. It uses the setup of model~1, but with some slight modifications. The electron
density and temperature in the interior cavity are now $\nel = 3 \times 10^4\,$cm$^{-3}$ and
$T = 2500\,$K, respectively. In the shell with the power-law profile, the temperature is $T = 2\times 10^4\,$K.
To demonstrate the maser amplification, we consider the emission from the shell, the cavity, and a
combination of both separately.

\begin{figure}
\centerline{\includegraphics[height=170pt]{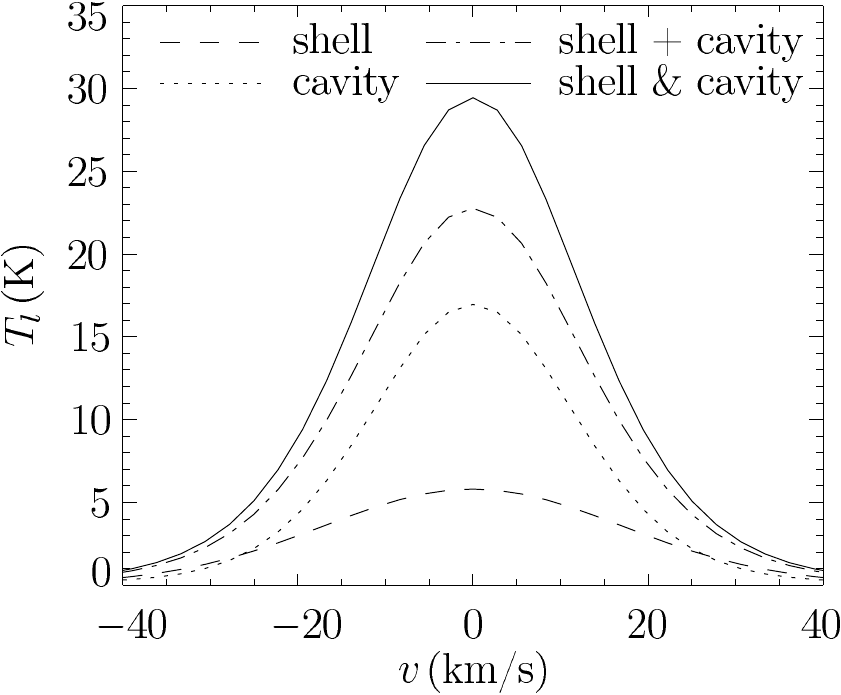}}
\caption{Source-integrated line temperatures of H85$\alpha$ for the shell and cavity components as well as a
combined model. The combined model exceeds the arithmetic sum of the components, demonstrating maser amplification.
Compare Figure 7 in \citet{vinvalhug79}.}
\label{fig:fig7}
\end{figure}

In Figure~\ref{fig:fig7} we show the source-integrated line temperatures of the H85$\alpha$ line.
The combined model that includes the shell and cavity simultaneously has a much larger line temperature than
the models with the shell or cavity alone. As the Figure demonstrates, the line temperature of the combined model
even significantly exceeds the arithmetic sum of the line temperatures from the model components. The agreement
with Figure~7 in \citet{vinvalhug79} is excellent.

\subsection{Adaptive-Mesh Test}
\label{sec:amrtest}

RADMC-3D has the capability of performing radiative transfer calculations on adaptive meshes. This is not
only highly demanded to post-process simulation data, but RADMC-3D can also perform on-the-fly
refinement on setups that are entirely defined by the user, such as in our test cases. We have already
applied this method in Section~\ref{sec:linstrwid} to better resolve the $r^{-2}$ singularity in model~1.
Here, we present additional tests of the adaptive mesh radiative transfer in RADMC-3D. We compare synthetic
maps for model~1 calculated on a homogeneous grid, on the Jeans-like refined grid presented in Section~\ref{sec:linstrwid},
and a grid in spherical coordinates with logarithmic radial grid spacings. The spherical grid has a resolution
in $(r,\vartheta,\varphi)$-coordinates of $100^3$. The cell boundaries $r_i$ of the radial collocation points are
chosen such that
\begin{equation}
r_i = r_\mathrm{in} \left(\frac{r_\mathrm{out}}{r_\mathrm{in}}\right)^{i / N_r}
\end{equation}
with the inner radius $r_\mathrm{in} = 10^{-3}\,$pc, the outer radius $r_\mathrm{out} = 0.1\,$pc and the number
of cells in radial direction $N_r = 100$. Here, the index $i$ runs from $0$ to $N_r$.

\begin{figure*}
\includegraphics[width=500pt]{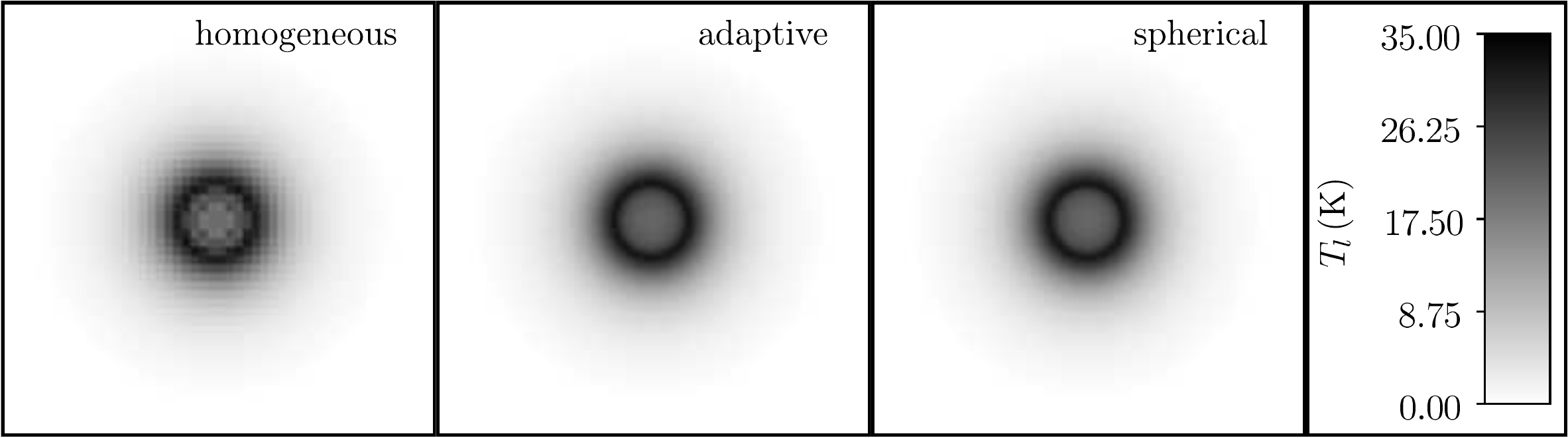}
\caption{Maps of the peak line temperature of the H137$\beta$ line for model~1. Shown are images for the
homogeneous grid (left), the adaptively refined grid (middle) and the spherically symmetric grid (right).}
\label{fig:amrtest}
\end{figure*}

Figure~\ref{fig:amrtest} displays the peak line temperature of the H137$\beta$ line for model~1.
As expected, the line temperature is the highest at the boundary of the cavity, where the electron
number density $\nel$ has its maximum. Of course, the spherical shape of the shell in model~1 is much
better represented in spherical coordinates. It is also cleary visible that the gas with high $\nel$
is better resolved on the adaptive mesh than on the homogeneous grid. Note, however, that the
numerical values of the line temperature agree very well for the three different grids.

\subsection{Relevance of Fine-Structure Splitting}

As already mentioned in Section~\ref{sec:physrecomb}, the recombination lines show fine-structure splitting
for frequencies above 100~GHz, which is the frequency range where ALMA operates. Thus, it is important to
know to which degree the fine-structure splitting affects the observational appearance of these lines.

Although the intensity distribution of the individual fine-structure components is highly asymmetric \citep{towleetal96},
the components are so close and the individual profiles are so strongly Doppler-broadened that deviations from
a Gaussian profile are not detectable, at least for typical \hii\ region temperatures around $10^4\,$K. As
\citet{towleetal96} show, the largest deviation of the brightest fine-structure component or of the intensity-weighted
mean to the Rydberg frequency can be at most of the order MHz for ALMA frequencies. However, our numerical
experiments show that the intensity of the superposition of all fine-structure components is systematically
lower than the intensity given by the \citet{menzel68} formula, and this difference can be of the order 10\%. Since the simultaneous
treatment of all fine-structure components is vastly more expensive than using a single line profile, we
choose to use the \citet{menzel68} approximation for ALMA frequencies but to keep in mind that the intensities
are slightly overestimated.

\end{document}